\documentclass[prl,nobalancelastpage,superscriptaddress,twocolumn,nofootinbib]{revtex4-2}

\usepackage{graphicx}
\usepackage{amsmath, amsxtra, amssymb, mathtools, nccmath, xspace}
\usepackage[utf8]{inputenc}
\usepackage{float}
\usepackage{placeins} 
\usepackage{xcolor} 
\usepackage{nicefrac} 
\usepackage{xspace}     
\usepackage[T1]{fontenc}

\newcommand{\bra}[1]{\langle #1|}
\newcommand{\ket}[1]{\left|#1\right\rangle}

\newcommand{\unit}{1\!\!1}

\begin{document}
\title{Entanglement transport and a nanophotonic interface for atoms in optical tweezers}

\newcommand{\Harvard}{Department of Physics, Harvard University, Cambridge, MA
02138, USA}
\newcommand{\Chicago}{Pritzker School of Molecular Engineering, University of Chicago, Chicago, IL 60637, USA}
\newcommand{\MIT}{Department of Physics and Research Laboratory of Electronics,
Massachusetts Institute of Technology, Cambridge, MA 02139, USA}

\author{Tamara \DJ{}or\dj{}evi\ifmmode \acute{c}\else \'{c}\fi{}}
\thanks{These authors contributed equally to this work}
\affiliation{\Harvard}
\author{Polnop Samutpraphoot}
\thanks{These authors contributed equally to this work}
\affiliation{\Harvard}
\author{Paloma L. Ocola}
\thanks{These authors contributed equally to this work}
\affiliation{\Harvard}
\author{Hannes Bernien}
\affiliation{\Chicago}
\author{Brandon Grinkemeyer}
\affiliation{\Harvard}
\author{Ivana Dimitrova}
\affiliation{\Harvard}
\author{Vladan Vuleti\ifmmode \acute{c}\else \'{c}\fi{}}
\affiliation{\MIT}
\author{Mikhail D. Lukin}
\thanks{To whom correspondence should be addressed; E-mail: lukin@physics.harvard.edu}
\affiliation{\Harvard}

\begin{abstract}
The realization of an efficient quantum optical interface for multi-qubit systems is an outstanding challenge in science and engineering. 	
Using two atoms in individually-controlled optical tweezers coupled to a nanofabricated photonic crystal cavity, we demonstrate entanglement generation, fast non-destructive readout, and full quantum control of atomic qubits.	
The entangled state is verified in free space after being transported away from the cavity by encoding the qubits into long-lived states and using dynamical decoupling. Our approach bridges quantum operations at an optical link and in free space with a coherent one-way transport, potentially enabling an integrated optical interface for atomic quantum processors.
\end{abstract}

\maketitle

Systems of neutral atoms have recently emerged as a powerful platform for quantum simulations and quantum computation~\cite{SaffmanReview10, Ebadi21,Scholl21,Madjarov20}. Atom trapping, coherent control, and readout are enabled by optical beam arrays, whereas high-fidelity multi-qubit quantum gate operations are enabled either through tunneling~\cite{Kaufman14}
or coherent excitation into atomic Rydberg states~\cite{Levine19}. 
Many potential applications of these systems, from quantum networking to modular quantum computing, require a mechanism for transferring qubit states to optical photons without compromising the ability for multi-qubit control~\cite{nickerson2013distibutedcomputing,monroe_interconnects}.
Integration with a quantum optical interface, such as an optical cavity with switchable interactions, would enable the required atom-photon entanglement generation and distribution~\cite{tiecke2014switch, northrup14,reiserer2015review}. A promising approach to achieve this is by spatially separating the cavity from the array and transporting selected atoms to and from the field as needed, preserving the coherence and scalability of atom array systems~(Fig. 1A).

We demonstrate a system with a micrometer-size photonic crystal cavity (PCC) coupled to neutral atoms in movable optical tweezers that can perform high-fidelity single qubit control and measurements as well as generate two-atom entanglement.
Following the prior work that demonstrated coupling of atoms to photonic crystal structures~\cite{Hood16,Chang_18,po20}, entangling atoms in macroscopic cavities~\cite{welte2017carving,welte2018gate}, and coherently transporting atoms and ions in free space~\cite{beugnon07,Lengwenus10, Pino2021}, we demonstrate PCC-mediated  entanglement generation that can be preserved and verified after transporting the atoms away from the cavity near-field.
This is achieved by engineering a compact high-cooperativity photonic structure, establishing full quantum control in the near-field, encoding the qubits into field-insensitive states, and using dynamical decoupling sequences synchronized with the atomic transport.

\begin{figure*}
    \centering
    \includegraphics[width=1.8\columnwidth]{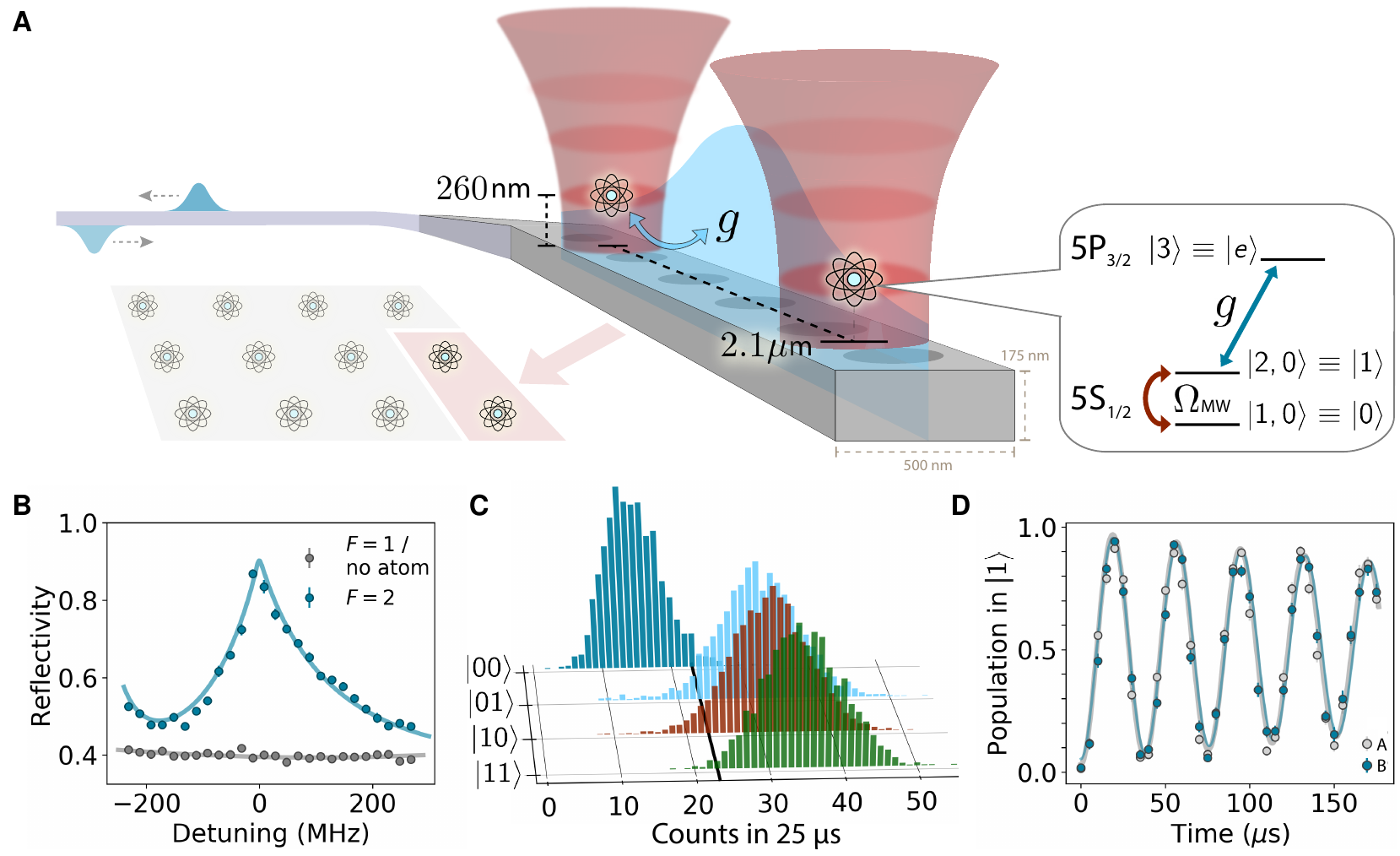}
    \caption{\textbf{Nanophotonic interface for atoms in tweezers.} 
\textbf{(A)} A nanophotonic cavity could act as a photonic link for atom arrays serving as quantum information processors. Our system consists of two atoms in optical tweezers coupled to a fiber-linked cavity capable of a coherent one-way transport. (Inset) Atomic qubit states coupled by a microwave drive $\Omega_{MW}$, with one coupled by the cavity field $g$ to an excited state. \textbf{(B)} Cavity reflection spectrum for an atom in a strongly coupled state (blue) and an uncoupled state (gray), described by ($2g$, $\gamma$, $\kappa$) = $2\pi \times$ (786, 6, 3800) MHz. 
\textbf{(C)} Readout of different two-qubit basis states by using the reflected photon counts. State $\ket{00}$ is distinguished from the others with fidelity $\mathcal{F} = 0.95(3)$. The atoms, cavity, and probe are on resonance. 
\textbf{(D)} Single-qubit Rabi oscillations in the cavity near-field showing preparation and coherent rotation with a $\pi$ pulse fidelity $\mathcal{F}_{\pi} = 0.946(3)$.}
    \label{fig1_main}
\end{figure*}

In our experiments, two ${}^{87}$Rb atoms were trapped in individually controlled optical tweezers and coupled to a PCC attached to a tapered optical fiber, enabling a low-loss photonic link out of the vacuum chamber~(Fig. 1A)~\cite{tiecke2015taperedfiber}. 
The small cross-sectional area of the PCC ($< 0.1~\mu$m$^2$) resulted in a small mode volume $V_m\sim 0.4 \lambda^3$ that enabled high atom-photon cooperativity~\cite{Chang_18} and allowed for switching of the photon interactions by moving the atoms over $\sim1\mu$m in and out of the cavity near-field.
Atoms at the PCC were trapped $260$~nm above the surface by a standing wave created with the retro-reflection of the tweezer~\cite{thompson2013couplingpc,tiecke2014switch,po20, si}. Loading into the first lattice site of the standing wave was accompanied by a trapping potential deformation that resulted in a $\sim50\%$ loading probability, which was increased to $80\%$ by performing polarization gradient cooling during the transport to the PCC~\cite{si}. (An efficiency above $90\%$ without cooling during transport has been achieved with atoms initialized in the motional ground state~\cite{thompson2013couplingpc}.)

The cavity reflectivity strongly depends on the internal atomic state, with qubits encoded in $|0\rangle \equiv |F=1, m_F = 0\rangle$ and $|1\rangle \equiv |F=2, m_F = 0\rangle$. The PCC was tuned to be resonant with the $5S_{1/2}, ~F=2 \rightarrow 5P_{3/2},~ F'=3$ optical transition and addresses all the $m_F$ levels within these manifolds owing to the large cavity width $\kappa$. The reflection spectrum of a single atom is characterized by the extracted cooperativity $C = 4 g^2/(\kappa \gamma) = 27(1)$ with a fit that accounts for thermal averaging (numbers in parentheses indicate one SD)~(Fig. 1B)~\cite{po20}. When probed on resonance, the difference in reflectivity between an atom in the coupled $F=2$ and uncoupled $F=1$ manifold was used to perform a non-destructive single-shot readout. With two atoms, the uncoupled state $\ket{00}$ is distinguished from the other two-qubit states with high fidelity~(Fig. 1C). This was used to determine the full two-atom state in all measurements at the PCC. For experiments in free space, readout was performed with state-selective push-out from the trap followed by fluorescence imaging. In both of these regimes, each pair of atoms was simultaneously prepared into $|0\rangle$ by using Raman-assisted state preparation~\cite{si}, and addressed with a global microwave drive to perform single-qubit gates~(Fig. 1D).

To coherently move the qubits from the near-field into free space, a millisecond coherence time was necessary. Choosing Zeeman-insensitive hyperfine sublevels as the qubit basis results in coherence times of $T_2^\prime = 2.1(1)$ ms and $9.7(8)$ ms in the near-field and free space, respectively, measured with a spin-echo pulse sequence~(Fig. 2A). In the field-insensitive basis, the coherence is typically limited by the thermal motion that samples varying differential ac Stark shifts in the trap~\cite{dephasing_meschede, dephasing_fiber}, and our measured $T_2^\prime$ in free space is consistent with this model, given the trap parameters. The $T_2^\prime$ at the PCC is likely further reduced by heating intrinsic to the near-field trapping geometry~\cite{hummer19, si}.

\begin{figure}
    \centering
    \includegraphics[width=0.8\columnwidth]{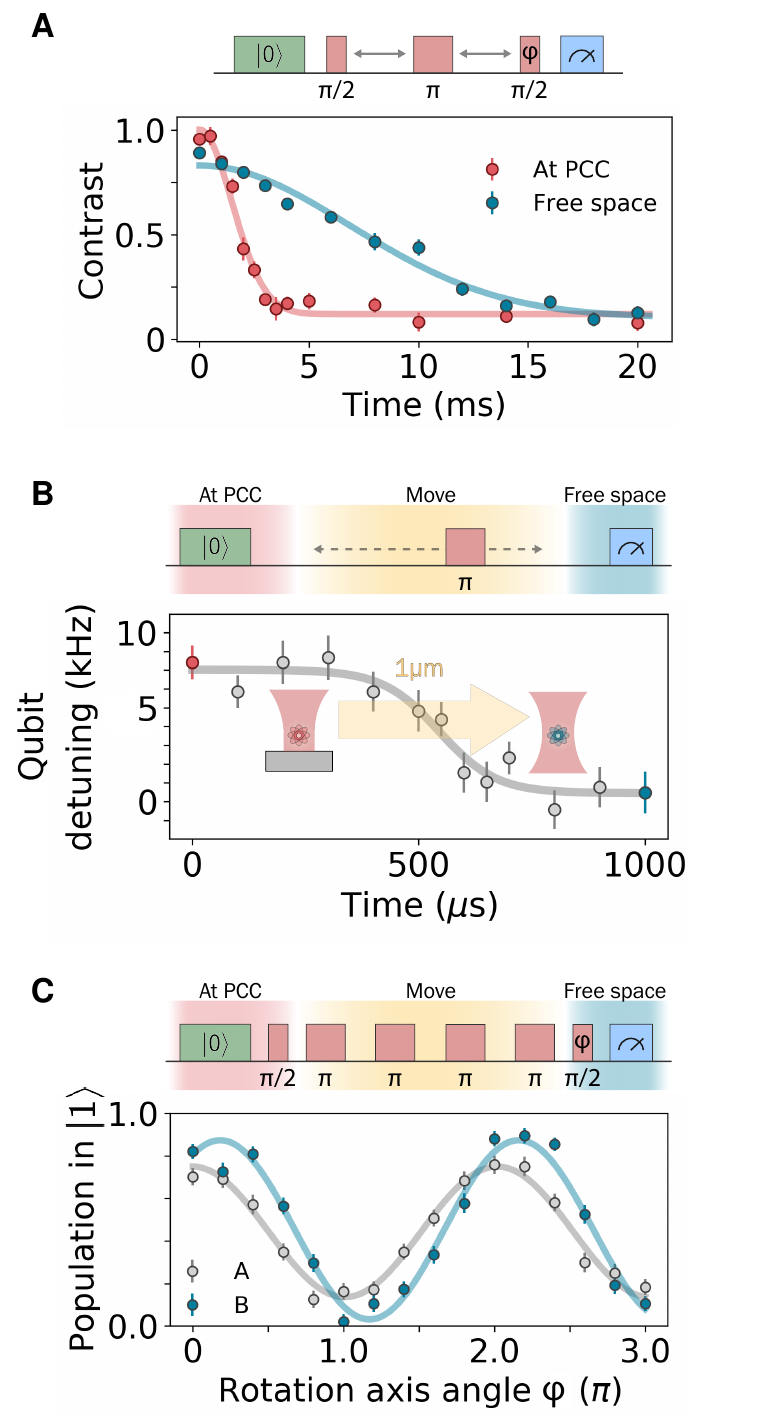}
    \caption{\textbf{Qubit coherence during the transport between the cavity and free space.}
\textbf{(A)} Measurement of the echo $T_2^\prime$ for stationary traps at the PCC (red) and in free space (blue). \textbf{(B)} Changes in the trapping potential configuration during transport cause differences in the light shift and therefore qubit frequency, which can lead to additional dephasing. The gray line indicates fit to the programmed ramp. \textbf{(C)} Single-atom coherence for atoms in each trap (labeled "A" and "B") after transport measured in a Ramsey-type sequence. A four-pulse Carr-Purcell sequence symmetric with respect to the qubit frequency profile during transport was used to echo out the light shift variations.}
    \label{fig2_main}
\end{figure}

Atomic coherence during transport is additionally affected by the change in differential ac Stark shift when moving between different trapping configurations.
To observe this change, a detuned microwave $\pi$ pulse was applied to single atoms at various times along the $\sim1~\mu$m path from the PCC, where the measured population transfer is proportional to the instantaneous qubit frequency~\cite{si}. From the resulting ramp profile in Fig. 2B, we obtained the minimal time of $650~\mu$s to move from the PCC, limited by the mechanical response bandwidth of the scanning mirrors.
Dephasing caused by fluctuations in the atom trajectory can be circumvented by using the Carr-Purcell decoupling pulse sequence~(Fig. 2C)~\cite{CP54}. This sequence was tested with the atom beginning in a superposition at the PCC and followed by $N \times \pi$ pulses along its transport, which performed better for an even $N$ because of the symmetry of the ramp profile shown in Fig. 2B. 
The performance of the decoupling sequence was characterized separately in the two optical tweezers (trap A and trap B): For an optimal $N=4$, an atom in trap B retains $0.88(5)$ of its initial coherence, limited by the move time and $T_2^\prime$~(Fig. 2C).
The lower contrast of $0.64(3)$ for an atom in trap A is ascribed to extra fluctuations in the scanning mirror, given that the traps are otherwise uniform.
In addition to decoherence, each atom has a $\sim15\%$ chance of being lost on the way out of the PCC, which is corrected for in the free space readout~\cite{si}.

\begin{figure*}
    \centering
    \includegraphics[width=1.8\columnwidth]{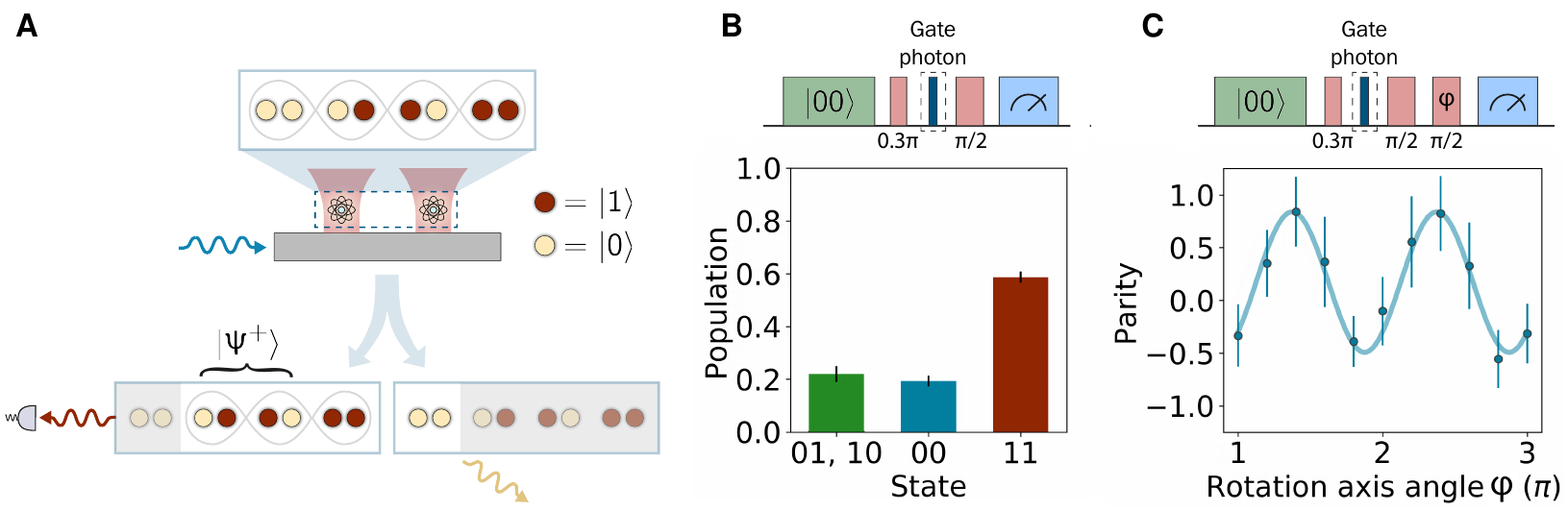}
\caption{\textbf{Heralded entanglement at the cavity.}
\textbf{(A)} Protocol description. Post-selection on a photon reflection from a superposition state carves out the $\ket{00}$ component, leaving the atoms predominantly in $\ket{\Psi^+}$. \textbf{(B)} Populations of the resulting state in different basis states. A $\pi/2$ pulse was applied before the measurement converting the Bell state $\ket{\Psi^+}$ into $\ket{\Phi^+}$. \{$P_{11}$, $P_{00}$\} is extracted \{with, without\} an additional global $\pi$ pulse before cavity reflection reflection. $P_{01}$ and $P_{10}$ cannot be distinguished. \textbf{(C)} Parity oscillations of the $\ket{\Phi^+}$ state after rotating over a varied axis $\phi$, showing non-classical correlations between
the qubits in all bases. The coherence $\rho_{11,00}$ is extracted from the amplitude. The combined coherence and populations gives a Bell state fidelity of $\mathcal{F} = 0.72(3)$. }
\label{fig3_main}
\end{figure*}

To generate an entangled state of atoms using the cavity,
we performed a protocol known as "cavity carving" that projectively measures a classical superposition state while maintaining qubit indistinguishability~\cite{ChenPRL15,Barontini15}, which can result in a maximally entangled Bell state~\cite{sorensen03,welte2017carving}. To implement the protocol, we initialized the atoms in $\ket{00}$, and rotated by an angle $\theta$ to create the state 
\begin{align}
    \ket{\Psi_0}  =  & \cos^2\left(\frac{\theta}{2}\right) \ket{00} -  \sin^2\left(\frac{\theta}{2}\right) \ket{11} \nonumber \\
   & - i \sin\left(\frac{\theta}{2}\right)\cos\left(\frac{\theta}{2}\right) \left( \ket{01}+\ket{10}\right)
\end{align}
We sent a weak coherent pulse into the cavity mode ($\overline{n} = 0.35$, chosen for negligible scattering error~\cite{si}), and condition on the detection of a reflected photon to project the state $\ket{00}$ out of the two-atom Hilbert space. The angle $\theta$ was chosen to maximize the overlap of the resulting state with $\ket{\Psi^+} = (\ket{01}+\ket{10})/\sqrt{2}$, minimizing the population in $\ket{11}$ while accounting for the finite reflectivity of $\ket{00}$. For our system, the $\ket{00}$ reflectivity is already less than that of the other two-atom states~(Fig. 1C). It was further reduced by an interferometer in the detection path and is ultimately limited by the interferometric stability~\cite{si}. After the projection, an additional $\pi/2$ pulse was applied to rotate the Bell state $\ket{\Psi^+}$ into $\ket{\Phi^+} = (\ket{00}+\ket{11})/\sqrt{2}$ for ease of characterization.

The entangled state was first characterized in situ by using the cavity reflection readout shown in Fig. 1C. 
The $\ket{\Phi^+}$ fidelity
$\mathcal{F}
=\bra{\Phi^+}\rho\ket{\Phi^+}
=\frac{1}{2}\big(
\rho_{00,00}+\rho_{11,11}\big)+
\operatorname{Re}(\rho_{00,11})
$ 
was determined by measuring populations and extracting coherence from the parity oscillation contrast upon another $\pi/2$ rotation over different axes~\cite{welte2017carving,si}. 
Populations are plotted in Fig. 3B with $\mathcal{P} = P_{00}+P_{11} = 0.78(3)$; the imbalance between $\ket{00}$ and $\ket{11}$ is discussed and modeled in the supplementary materials~\cite{si}. 
The parity oscillation in Fig. 3C corresponds to a coherence of $\rho_{11,00} = 0.33(2)$.
These measurements give a fidelity of $\mathcal{F} = 0.72(3)$ and a concurrence bound of $\mathcal{C} \geq 0.46(5)$~\cite{si}. The major part of the error~($24\%$) is accounted for by the chosen initial angle $\theta$ for our protocol, with an additional $4\%$ from the initialization imperfections. Although our approach uses a single gate photon, heralding on repeated gates can improve the entanglement fidelity at the expense of the success probability~\cite{welte2017carving}.

Last, the entangled atoms were transported away from the cavity while applying decoupling pulses. The qubit state populations were measured in the $ZZ$, $XX$, and $YY$ bases~(Fig. 4), giving the population $\mathcal{P} = 0.78(6)$ and the coherence $\rho_{11,00} = 0.26(5)$, corrected for the atomic loss. This resulted in a fidelity of $\mathcal{F} = 0.65(6)$ and a lower bound to the concurrence of $\mathcal{C} \geq 0.35(11)$ ($\mathcal{F}_u = 0.52(5)$ and $\mathcal{C}_u \geq 0.04(10)$ without the readout correction). Both $\mathcal{F} > 0.5$ and $\mathcal{C} > 0$ certify entanglement after transport at a $99\%$ confidence level. The state fidelity is consistent with the decoherence observed for single atoms during transport (Fig. 2C), limited by atomic thermal motion and uncorrelated fluctuations in the programmed trajectories. The trajectory fluctuations can be reduced in the future by using an acousto-optic deflector (AOD) that would speed up the transport and enable atomic displacements of more than 100 $\mu$m, which is sufficient for high-fidelity Rydberg state control~\cite{Kubler2010}.

\begin{figure}
    \centering
    \includegraphics[width=0.9\columnwidth]{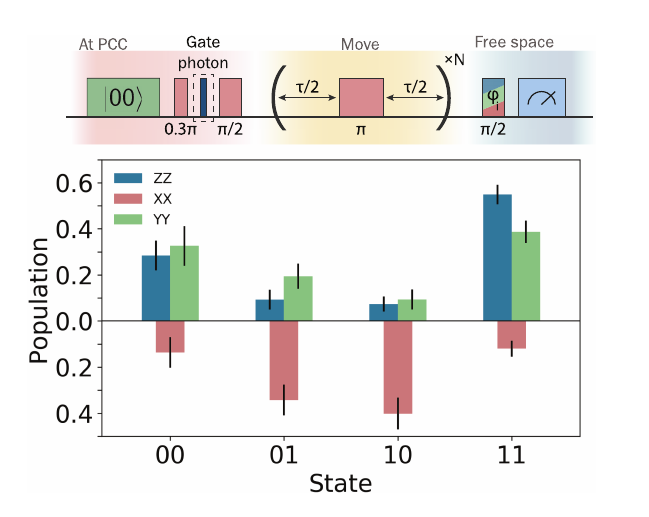}
\caption{\textbf{Coherent transport of entangled atoms from the cavity to free space.}
After entangling the atoms at the cavity, the tweezers are moved away while applying dynamical decoupling. Free space readout in different bases after transport shows the remaining entanglement fidelity of $\mathcal{F} = 0.65(6)$, which is larger than the classical limit of $0.5$. } 
    \label{fig4_main}
\end{figure}

These results demonstrate that combining movable atoms in tweezers with nanophotonics constitutes a promising platform for the development of a quantum optical node for large-scale atomic quantum processors. If integrated with a free space atom array, our system could enable quantum teleportation of the atom array states to optical photons~\cite{Bennett93} or remote entanglement of arrays. This can be accomplished by using the demonstrated one-way transport, in which the atoms that are entangled with a photon at the PCC are then coherently moved to the free space array to interact through Rydberg gates. Extending our demonstration to a general-purpose quantum optical interface with a coherent two-way transport could enable feed-forward protocols such as quantum error correction by means of fast, high-fidelity quantum nondemolition cavity readout. Furthermore, with multiple atoms coupled to nanophotonic waveguides, entangled and arranged into arbitrary spatial configurations, important classes of many-particle quantum states can be created, such as atomic cluster or graph states~\cite{Choi19_cluster_st}, large-scale photonic Fock states~\cite{Paulisch19}, or Schr{\"o}dinger cat states~\cite{Davis18}. Last, telecom transitions present in both alkaline-earth and alkali atoms~\cite{Covey19,Menon20} could extend atom-nanophotonic quantum networking to long-distances, which is crucial for quantum communication and the scale-up of processing power~\cite{monroe_interconnects}.

\section*{Acknowledgments}
We thank
Crystal Senko,
Harry Levine,
Dolev Bluvstein,
Ralf Riedinger,
Mihir Bhaskar,
David Levonian, 
Johannes Borregaard,
Florentin Reiter,
Sylvain Schwartz,
Jeff Thompson,
and Alexander Zibrov
for useful discussions and experimental contributions. 
\textbf{Funding:}
This work was supported by the Center for Ultracold Atoms, the National Science Foundation, Air Force Office of Scientific Research Multidisciplinary University Research Initiative, Vannevar Bush Faculty Fellowship, US Department of Energy (DOE) QSA Center (contract 7568717), and Army Research Laboratory CDQI. B.G. acknowledges support from the US Department of Defense (DOD) NDSEG. The device was fabricated at the Harvard Center for Nanoscale Systems (CNS) (NSF ECCS-1541959).
\textbf{Author contributions:} 
T.\DJ., P.S., P.L.O., and H.B. built the experimental setup, performed the measurements, and analyzed data. B.G. and I.D. assisted with experiments. 
All work was supervised by V.V. and M.D.L. All authors 
discussed the results and contributed to the manuscript. 
\textbf{Competing interests:} 
V.V. and M.D.L.  are co-founders and shareholders of QuEra Computing. The other authors declare no competing interests. 
\textbf{Data and materials availability:}  All data needed to evaluate the conclusions in the
paper are present in the paper and the supplementary materials.

\FloatBarrier

\bibliographystyle{Science}

\clearpage
\newpage

\renewcommand{\thefigure}{S\arabic{figure}}
\renewcommand{\thetable}{S\arabic{table}}
\setcounter{figure}{0}
\setcounter{equation}{0}

{
\renewcommand{\addcontentsline}[3]{}
\section{Supplementary Materials}
}
\section{Cooperativity estimate}

The reflectivity amplitude of an atom-cavity system driven at a frequency $\omega$ can be calculated as \cite{po20}:

\begin{equation}
    r = \kappa_{wg} \left( \frac{\kappa}{2} - i\Delta_c + \frac{g^2}{\frac{\gamma}{2}-i \Delta_a} \right)^{-1} - 1
    \label{eqn:r}
\end{equation}
where $\gamma$ is the atomic spontaneous decay rate, $g$ is the single-photon Rabi frequency,  $\Delta_a = \omega - \omega_a$ is the atomic detuning, and $\Delta_c = \omega - \omega_c$ is the cavity detuning. The cavity decay rate $\kappa$ is the sum of decay into the waveguide  $\kappa_{wg}$, and the decay elsewhere at $\kappa_{sc}$. The single-atom cooperativity is defined as $C = 4g^2/(\kappa \gamma)$. The measured reflection spectra correspond to $|r(\omega)|^2$. The entanglement protocol implemented in this work is done with the atom, cavity, and probe photons on resonance, in which case the reflectivity simplifies to
\begin{equation}
    R = |r|^2 = \left| 1 - \frac{2\kappa_{wg}}{\kappa} \frac{1}{C} \right|^2
    \label{eqn:refl}
\end{equation}
showing a highly nonlinear dependence on cooperativity. For our measured cooperativity, the atom-cavity system acts as a near-perfect mirror already with a single coupled atom.

The reflection spectrum in Fig.~1B appears asymmetric due to the coupling to $F'=1$ and $F'=2$ excited states in addition to the main $F'=3$ and is fitted using the theoretical model described in \cite{po20}. According to this model, the thermal motion of a trapped atom next to the cavity results in a distribution of cooperativities sampled by the atom. Using the independently measured radial trap frequency of $2\pi \times 115$ kHz, axial trap frequency of $2\pi \times 550 $ kHz, and final temperature of $T = 70$ $\mu$K, we fit for the cooperativity an atom would experience if it were stationary and obtain $C_0 = 47 (1)$. The distribution of the cooperativities an individual atom samples has a mean value of $27(1)$, and a standard deviation of $25(2)$. We adjust the two trap positions relative to the cavity mode to match their individual mean cooperativities while not overlapping their traps, resulting in a $2.1 $ $\mu$m distance between the atoms centered around the mode maximum. At the mode maximum of the cavity the mean single-atom cooperativity is measured to be 71(4) \cite{po20}.

\subsection{Experimental sequence}
\begin{figure*}
\centering
\includegraphics[width=1.8\columnwidth]{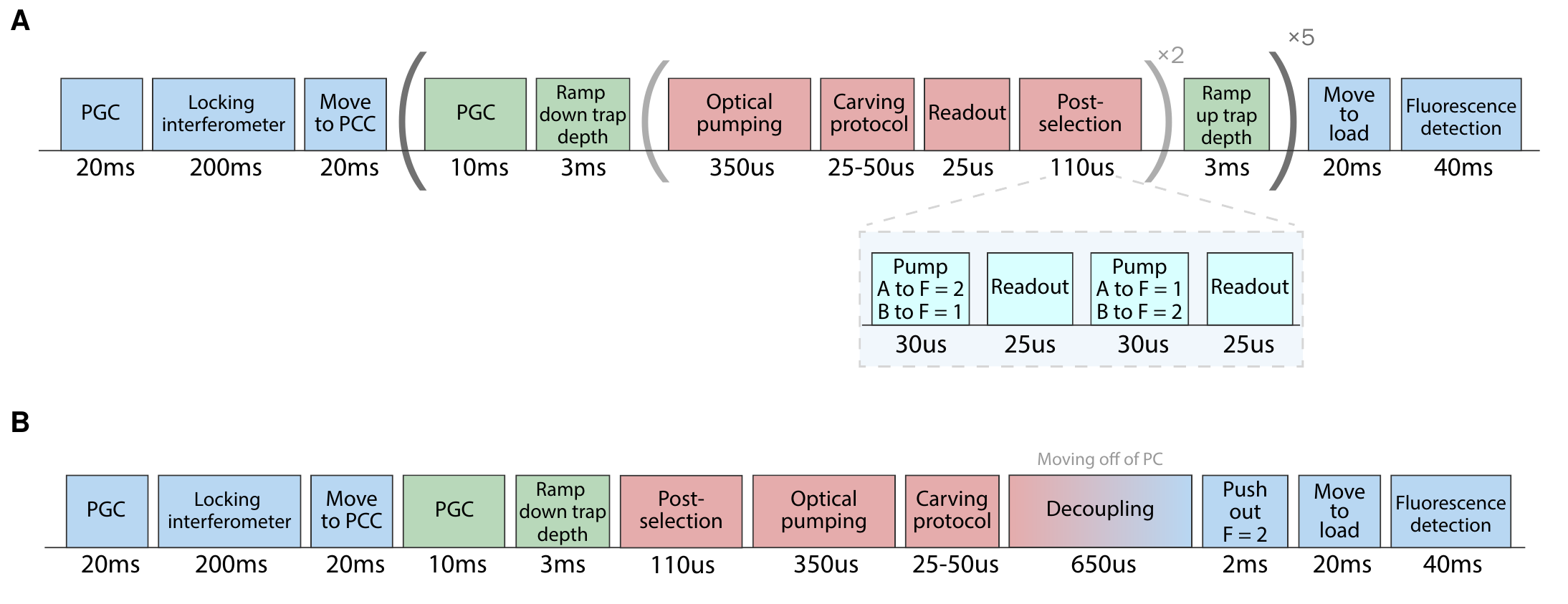}
\caption{ Experimental sequence.
\textbf{(A)} Time breakdown of a typical two-atom entanglement sequence between atom-A and atom-B. 
\textbf{(B)} An experimental sequence for the coherent transport of the atomic states that incorporates phases both at the cavity (red and green) and in free space (blue). }
\label{fig:exp_time}
\end{figure*}

The total run time can be separated into periods of experimental run ($\sim 85 \%$ of the total time), equipment programming, and setup calibration. Confocal tracking of the cavity position accounts for most of the setup calibration and is needed to consistently move atoms to the same position in the cavity mode. This is achieved by imaging the cavity in different focal planes for 2 minutes every 25 minutes and recalibrating the coordinates for each dipole trap. The cavity resonance is tuned by heating the cavity with an off-resonant infrared beam, thus without introducing additional elements or electric fields near the atoms. The lock provides active feedback throughout the experiment but can be frozen or jumped to a different cavity detuning value if needed.   

A typical experimental sequence is outlined in Fig. \ref{fig:exp_time}. We load atoms from a magneto-optical trap with the tweezers positioned $50$ $\mu$m from the cavity \cite{po20}. Each experimental trial begins upon detection of atomic fluorescence indicating presence of both atoms. At this position, we perform polarization gradient cooling (PGC) and bring the temperature of the atoms down to 15 $\mu$K. The dipole traps are subsequently moved to the cavity with galvanometer mirrors. 
During transport, the trap transforms from a Gaussian beam into a 1D optical lattice with tight axial confinement due to its reflection from the PCC. The increase in the trap frequencies results in a higher atomic temperature. 
To counteract this heating, we apply PGC again while attempting to load the atoms into the first lattice site closest to the PCC. We achieve 80$\%$ single atom loading probability with a final temperature of $70$ $\mu$K, up from a 50$\%$ loading probability without cooling during loading. In the case with cooling, we have determined that the remaining 20$\%$ consists of approximately 15$\%$ loss and 5$\%$ higher lattice site loading. These values were characterized by combining a measurement evaluating the two-way survival of an atom, a reflection measurement determining the presence of an atom in the first lattice site immediately after loading, and the survival of an atom moved to free space after confirming its presence in the first lattice site. While at the cavity, the atoms experience a higher heating rate than in free space, reducing their trap lifetime to $180$ ms. Photon scattering due to state readout through the cavity causes additional heating, limiting the number of experimental repetitions at the cavity.  Furthermore, to reduce light shift-induced inhomogeneities, we reduce the trap depth from $2\pi\times50$~MHz to $2\pi\times32$~MHz before preparing the qubit state (see Section 2.1 below), which increases the loss probability. For the experiments where the atoms are kept next to the PCC the entire time (such as in Figs. 1 and 3 in the main text), we re-cool the atoms every two repetitions, and repeat the experiment up to 10 times, as shown in Fig. \ref{fig:exp_time}A. For a typical experiment, two atoms are on average present in 4 of those iterations. For experiments which incorporate phases both at the cavity and in free space (such as in Figs. 2 and 4 in the main text), we do not re-use the same atom pair (Fig. \ref{fig:exp_time}B).       

Since the atoms can be heated out of the traps during the experimental attempts, we postselect the experiments based on atomic presence detected via the cavity. As shown in Fig. 1C, the cavity reflection cannot distinguish the number of coupled atoms. By adding individual addressing beams co-propagating with the dipole traps and resonant with the $5S_{1/2}\rightarrow 5P_{1/2}$ transition, we can selectively pump either atom to the coupled state ($F = 2$) while leaving the other in the uncoupled state ($F = 1$). This allows us to perform entanglement experiments where both atoms are loaded with >$99\%$ and to overcome the common problem of probabilistic loading from which all nanophotonic platforms suffer \cite{goban2015superradiance,kim_clhung2019chiptrapping,Schneeweiss2020}.

Combining the probabilistic loading at the cavity, the two-atom initial trigger rate, and the single photon detection rate for the entanglement experiments gives a Bell pair generation rate of $0.5$ events/min. The largest contribution to the low data rate is the number of collected photons from the coupled level during the gate pulse. The total number of collected photons is equal to $N_{collected} = T_{int} \eta N_{sent} = 10^{-2}$, where $\eta = 0.28$ is the total detection efficiency, $N_{sent} = 0.35$ is the number of photons sent to the cavity, and $T_{int} = 0.1$ is the throughput from the interferometer in the undercoupled cavity regime. We discuss the interferometer throughput $T_{int}$ in detail in the section about the entanglement scheme implementation. The total detection efficiency $\eta$ consists of the single-photon counter (PerkinElmer SPCM-AQR-16-FC) quantum efficiency ($0.6$), fiber taper coupling efficiency ($0.6$), and optical path throughput ($0.8$), and can be readily improved with superconducting nanowire detectors and a higher PCC-fiber coupling efficiency.

\subsection{State preparation}

\begin{figure}
\centering
\includegraphics[width=0.65\columnwidth]{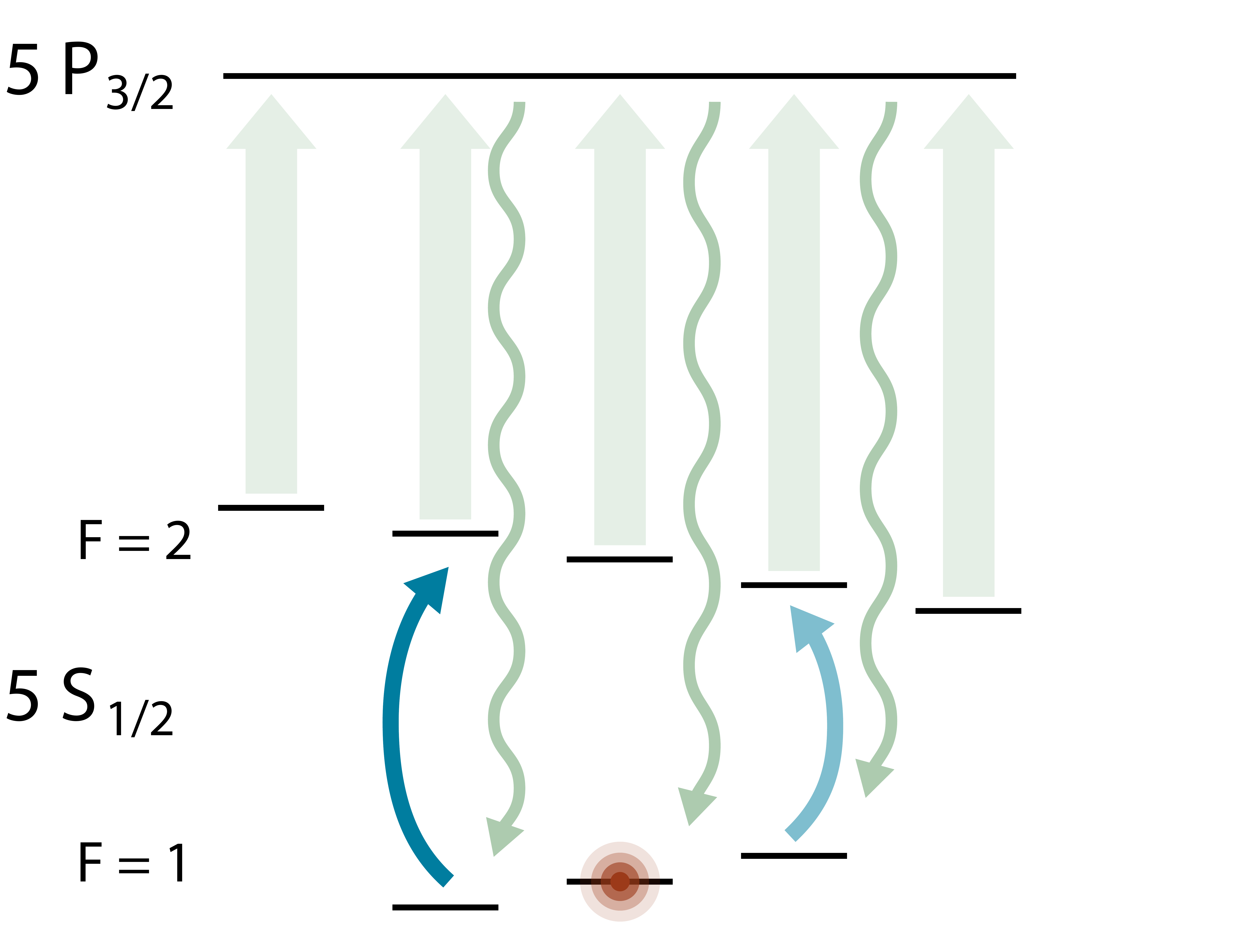}
\caption{An illustration of the Raman-assisted state preparation. Cycling between repumping the atom into $F = 1$ (green) and depleting the population in $m_F = \{-1, ~1\}$ (blue) leaves the atom in $\ket{1,0}$. }
\label{fig:op}
\end{figure}

To prepare an atom in the qubit state $\ket{F,m_{F}} = \ket{1,0} \equiv \ket{0}$ we perform a frequency-selective Raman-assisted optical pumping protocol. The protocol consists of sequentially pumping the atomic population into the $F=1$ ground state manifold followed by driving $\ket{1,\pm 1} \rightarrow \ket{2,\pm 1}$, leaving the target qubit state dark. The latter is accomplished through driving Raman transitions between the two ground state hyperfine manifolds with a circularly polarized beam. This beam is 100 GHz detuned from the $5S_{1/2} \rightarrow 5P_{1/2}$ transition and contains two frequency components detuned by the hyperfine splitting of 6.8 GHz generated by an amplitude modulator (Jenoptik AM785). It co-propagates with one of the dipole traps through the objective and is large enough to equally address both atoms which are separated by 2.1 $\mu$m. We begin the optical pumping cycle with the atomic population coarsely pumped to $F=1$, prepared with the MOT beams tuned to the $F=2 \rightarrow F'=2$ transition. 
The Raman beat note is generated from the microwave source (SRS SG 384) operated in the frequency-shift keying (FSK) mode which allows rapid frequency switching to sequentially drive the $\ket{1,\pm 1} \rightarrow \ket{2,\pm 1}$ ground state transitions, thus leaving the final population in $\ket{1,0}$ (Fig. \ref{fig:op}). Afterwards, we apply the global MOT beams again to pump the population from $F=2$ back to $F=1$. We repeat this cycle 40 times, with each Raman pulse taking 2 $\mu$s and the MOT pumping pulse taking 2.2 $\mu$s. The optical pumping sequence takes 340 $\mu$s in total and results in a $0.98(2)$ preparation fidelity in $\ket{1,0}$. The optical pumping fidelity estimation is extracted via the Rabi oscillation contrast (Fig. 1D), where the $\ket{1,0} \rightarrow \ket{2,0}$ transition is driven and the resulting $F=2$ population is read out with the cavity. This optical pumping fidelity is corrected for the cavity readout infidelity. 
Our optical pumping scheme overcomes the challenge of maintaining polarization purity near the PCC, where the atom samples near-field polarization variation of an optical field reflected from the nanostructure. This makes other polarization-based optical pumping schemes inapplicable for high-fidelity state preparation. Our scheme, however, relies only on frequency selectivity between the different magnetic sublevels, enabled by a Zeeman shift of $12$ MHz ($B = 17 $~Gauss). 

The state preparation is performed on both atoms simultaneously, while they are strongly coupled to the PCC. The resonant cavity picks up stray light very efficiently, so we pay special attention to shuttering off all the unused light sources. Other than that, we observe no adverse effects of the strong coupling on the preparation fidelity. Since the state preparation scheme uses only optical sources, it can be adapted so that it prepares each atom individually, without disturbing the atoms in the surrounding tweezers, which may be useful for the proposed integration with a free-space atom array.

\subsection{Two-atom readout in free space}

The experiments described in the main text performed with tweezers positioned away from the PCC
rely on the standard push-out state detection.
We apply a pulse resonant with the
$F=2 \rightarrow F'=3$ transition that removes atoms in $F=2$ and record
fluorescence to detect the atoms remaining in $F=1$.
The push-out pulses and fluorescence detection
are performed $\sim1~\mu$m and 50~$\mu$m away from the PCC respectively.
We characterize the readout contrast by preparing the atoms in the
$F=\{1,2\}$ manifolds at the PCC, moving the tweezers away at a trap depth of $2\pi\times 32 $~MHz, and performing the push-out readout,
obtaining the retention probability of $\sim \{0.80,0.05\}$. 
We recalibrate the retention levels every half an hour in parallel with running the entanglement experiments. 
The finite retention of $F=2$ is limited by imperfections in the push-out procedure.
The retention of $F=1$ is primarily limited by transport loss ($\sim15\%$) and loss from collisions with the MOT atoms during the imaging ($\sim5\%$). The transport loss depends on the trap depth and initial temperature, and for our parameters can be suppressed at a trap depth of $2\pi\times 50 $~MHz.

As the finite contrast of the push-out detection limits the single-shot detection fidelity,
we infer the population components from the readout probability after
collecting meaningful statistics based on the readout calibration described above. In the case of two-atom experiments, the
probabilities of retention $\mathcal{R}$ are given by
\begin{equation}
  \mathcal{R}
  =
  \begin{pmatrix}
  R_{00}\\R_{01}\\R_{10}\\R_{11}
  \end{pmatrix}
  = \mathcal{K}
  \begin{pmatrix}
    P_{00}\\P_{01}\\P_{10}\\P_{11}
  \end{pmatrix},
\end{equation}
where $R_{ij}$ denotes the results corresponding to the four possibilities of
retention of the two atoms, $P_{ij}$ are the populations in the corresponding two-atom states,
and $\mathcal{K}$ is a correction matrix given by

{\begin{align}
  &\mathcal{K} =
  \mathcal{K_A} \otimes \mathcal{K_B} \nonumber = \\ &
  \begin{pmatrix}
    h_{A}h_{B} & h_{A}(1-h_{B}) & (1-h_A)h_B & (1-h_{A})(1-h_{B})\\
    h_{A}l_{B} & h_{A}(1-l_{B}) & (1-h_A)l_B & (1-h_{A})(1-l_{B})\\
    l_{A}h_{B} & l_{A}(1-h_{B}) & (1-l_A)h_B & (1-l_{A})(1-h_{B})\\
    l_{A}l_{B} & l_{A}(1-l_{B}) & (1-l_A)l_B & (1-l_{A})(1-l_{B})
  \end{pmatrix}^\intercal
\end{align}}
where $\{h_i,l_i\}$ denote the retention probabilities of atom $i$ prepared in $F=\{1,2\}$.
For a generic two-atom measurement, we multiply the inverse of the correction matrix
with the measurement result to obtain the populations \cite{Bernien13}.

\subsection{Readout through the cavity} 
\begin{table}[t]
\begin{center}
    \begin{tabular}{c | c | c}
  State & Mean  & Confidence interval \\
  \hline
  \hline
   $\ket{00}$ &  $0.956$ & $  [0.936, 0.975] $\\
  \hline
  $\ket{01}$ &  $0.949$ & $  [0.922, 0.964] $\\
  \hline
  $\ket{10}$ &  $0.948$ & $  [0.921, 0.961] $\\
  \hline
  $\ket{11}$ &  $0.991$ & $  [0.987, 0.996] $\\
  \hline
  
\end{tabular}
\end{center}
  \caption{Readout fidelities of different two-atom states}
  \label{table:ReadoutFidelities}
\end{table}

The measurements in Fig. 1 and 3 in the main text were done by reading the atomic states using the cavity reflection. As shown in the Fig. 1C, readout through the PCC distinguishes the state $\ket{00}$ from the rest of the states in the two qubit manifold. By integrating for $25~\mu$s and choosing a photon count threshold, we perform a single shot readout which distinguishes $\ket{00}$ from the other states with the fidelities reported in Table \ref{table:ReadoutFidelities}. The confidence intervals for the fidelities were obtained by changing the threshold by $\pm1$ photon.

\subsection{Fidelity extraction from parity}
To characterize the state after the carving protocol and its overlap with $\ket{\Phi^{+}}$, we perform two sets of measurements. In the first we measure the populations $P_{00}$ and $P_{11}$ and in the second we measure the parity oscillations of the state $\ket{\Phi^{+}}$. The populations in $\ket{00}$ and $\ket{11}$ are extracted by reading out the states without and with a global $\pi$ pulse applied before the cavity reflection detection respectively. Using the normalization condition $\sum P_{ij} = 1$, we also extract the population in the odd states, $P_{odd} = P_{01}+P_{10} = 1 - (P_{00}+P_{11})$. We can therefore measure the parity of the state, defined as 
\begin{equation}
    \Pi  \equiv P_{00} - P_{01} - P_{10} +P_{11} = 2(P_{00}+P_{11})-1
    \label{eqn:parity}
\end{equation}

We calculate the fidelity of preparing the Bell state $\ket{\Phi^{+}}$ as $\mathcal{F}
=\bra{\Phi^+}\rho\ket{\Phi^+}
=\frac{1}{2}\big(
\rho_{00,00}+\rho_{11,11}+\rho_{00,11}+\rho_{11,00}
\big)$ 
by measuring the populations $\{\rho_{00,00},\rho_{11,11}\}$ and the coherences
$\{\rho_{11,00}=\rho_{00,11}^*\}$.
The total population is directly measured from the cavity readout. The coherence is recorded after a $\pi/2$ rotation over a variable axis $\phi$ \cite{monroe2000,welte2017carving}
and is extracted from the amplitude of the parity oscillation as

\begin{align}
  \Pi (\phi) = ~& 2 \operatorname{Re} (\rho_{10,01}) + 2 \operatorname{Im} (\rho_{11,00})\sin(2\phi) \nonumber \\
  &+ 2 \operatorname{Re} (\rho_{11,00})\cos(2\phi)
\end{align} 
Combining these two measurements gives the reported fidelity of $\mathcal{F} = 0.72(3)$. Note that the data in Fig. 3 and the reported entanglement fidelity at the PCC are not corrected for the finite readout fidelity. 

The fidelity measurement after the transport is done in an analogous way. All four populations $P_{ij}$ are extracted from the free space fluorescence detection. 
The phase of the coherence term $2\phi_0 = \arg(\rho_{11,00})$ is calibrated using each individual atom's Ramsey fringes, after which the resulting two-atom state coherence is extracted from the difference in the parity measurements $\Pi (\pi/2) - \Pi (0)$ between the bases \{parallel, orthogonal\} to $\phi_0$, labeled \{$XX$, $YY$\} (Fig.~4B). 
Combining this measurement with the populations ($ZZ$ basis) gives the measured fidelity of $\mathcal{F} = 0.65(5)$.

\subsection{Concurrence bound}

Concurrence is a basis-independent measure of how entangled a state is and is therefore a more general entanglement measure than the fidelity of preparing a certain state. In practice, however, measuring the concurrence requires knowledge of the full density matrix of the prepared state. For this reason, a lower bound is often used to check if a state is entangled \cite{wootters97}. A concurrence of $C>0$ proves entanglement.

Given the amplitude of the parity oscillations we obtain a lower bound on the concurrence, following \cite{Nguyen19_prb}. In the ZZ basis, we primarily prepare $|\Phi^+\rangle =  \left( |00\rangle + |11\rangle \right)/\sqrt{2}$. 
The lower bound for the concurrence is then
\begin{equation} 
\mathcal{C}\geq 2 (|\rho_{00,11}|-\sqrt{\rho_{01,01}\rho_{10,10}})
\label{eqn:concurrence}
\end{equation}

Using the Eqn. \ref{eqn:concurrence}, we directly extract concurrence after moving of $\mathcal{C}\geq 0.35(11)$. At the cavity we only measure the sum of the two populations $P_{01}+P_{10} = 0.22 \pm 0.03$. To get a lower bound, we want the maximum of their product, which occurs when they are equal $P_{01}=P_{10}$. Using this, we extract $\mathcal{C}\geq 0.46(5)$.

\subsection{Entanglement scheme using the interferometer}

\begin{figure}
\centering
\includegraphics[width=0.7\columnwidth]{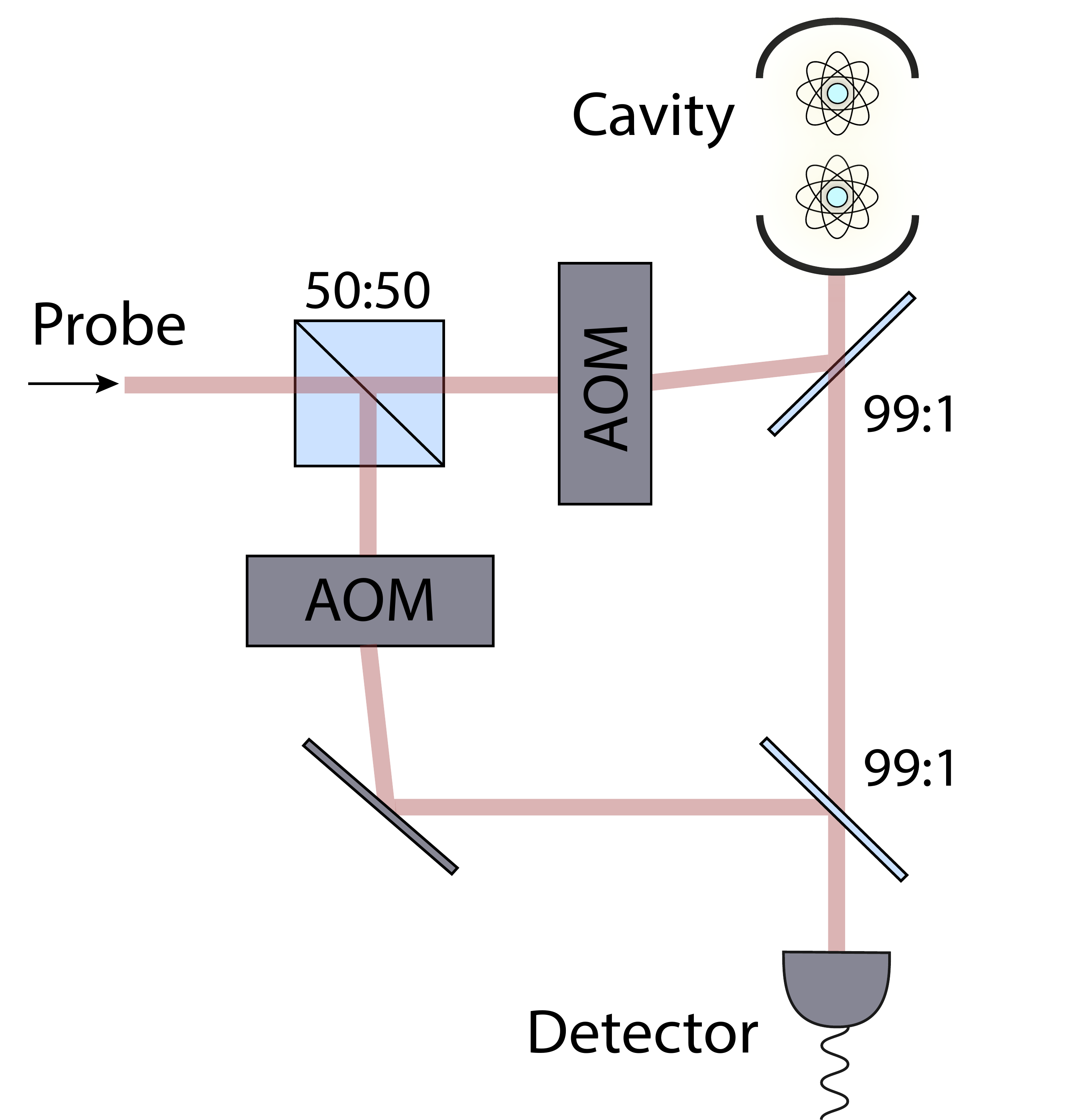}
\caption{Sketch of the interferometer in the detection path used to improve the reflection contrast between a cavity with and without coupled atoms.}
\label{fig:interf}
\end{figure}

The entanglement fidelity of the coherent photon reflection protocol is determined by the reflection amplitudes of different two-atom basis states. In turn, the reflection amplitudes determine the optimal initial rotation angle $\theta$. The four qubit states have reflectivities given by Eqn. \ref{eqn:refl}, which for our parameters are $R_{00} = 0.40$, $R_{01} = R_{10} = 0.94$, $R_{11} = 0.97$. Since the entanglement protocol relies on distinguishing $\ket{01}$, $\ket{10}$ and $\ket{11}$ from $\ket{00}$ as well as possible, we add an interferometer in the detection path to eliminate the uncoupled reflectivity $R_{00}$. The cancellation is achieved using a Mach-Zehnder interferometer with the relative phase tuned for destructive interference of the uncoupled reflectivity (Fig. \ref{fig:interf}). In this way, any cavity with a finite reflectivity on resonance can be used to implement entanglement via photon carving. The interferometer can be switched on and off during an experiment by changing the AOM driving amplitudes to allow for both the single photon gating (interferometer on) and atomic state detection (interferometer off). 

\begin{figure*}
\centering
\includegraphics[width=1.4\columnwidth]{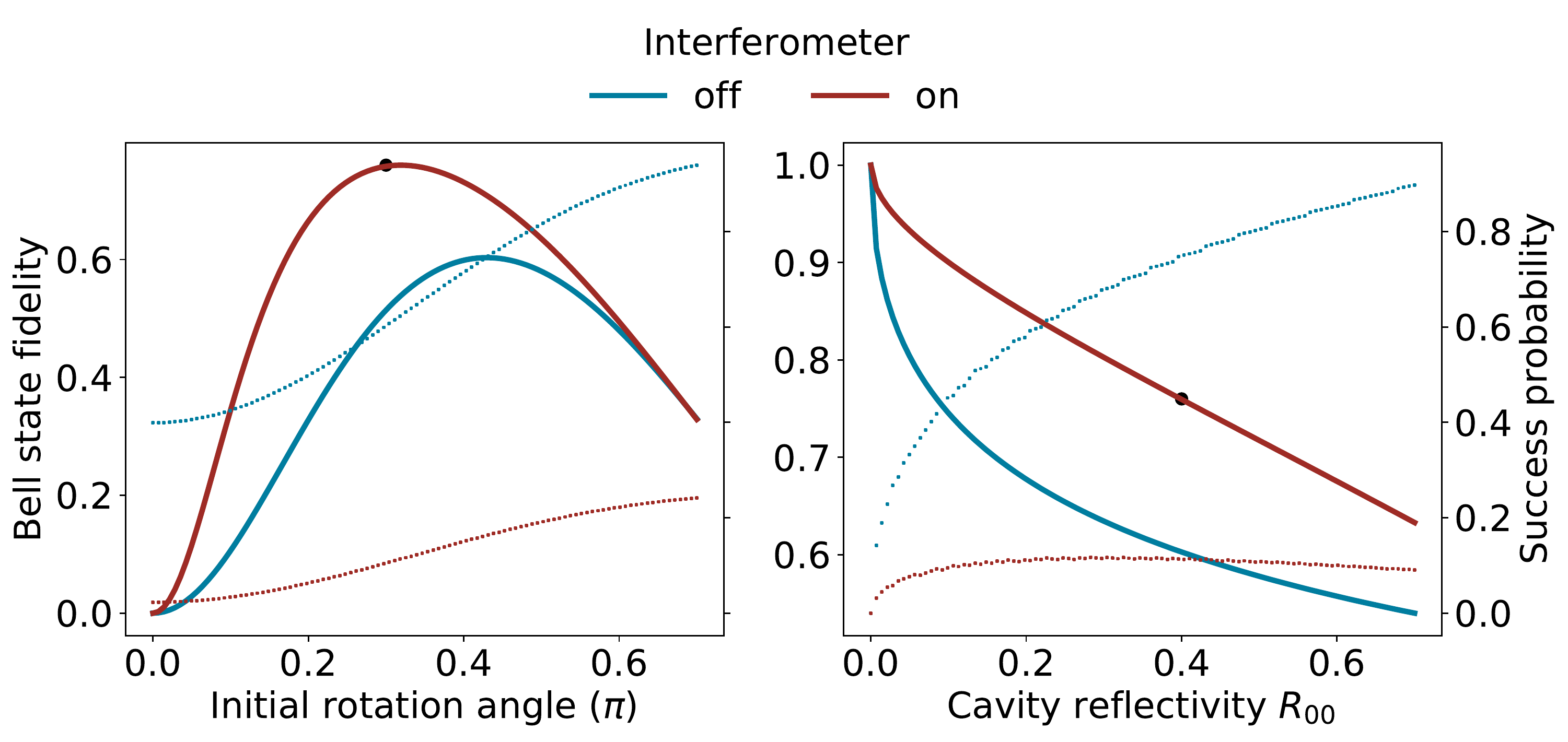}
\caption{Parameters that determine the entanglement fidelity. (Left) Dependence of the maximal Bell state fidelity (solid lines, left y axis) and protocol success probability (dotted lines, right y axis) on the initial angle and interferometer contrast. Not using the interferometer would limit the maximal fidelity to $\sim0.6$. With our measured interferometer contrast of $0.96$, the fidelity is limited to $0.76$ (black dot). (Right) Keeping the interferometer contrast the same, the protocol maximal fidelity can be improved by using a cavity with a smaller reflectivity $R_{00}$. }
\label{fig1:angle}
\end{figure*}

In the case of a perfect interferometer, $R_{00} \rightarrow 0$ and would yield a fidelity of $\mathcal{F} = 1/(1+\tan(\theta/2))^2$, which in principle can be $\mathcal{F} \rightarrow 1$ with an infinitesimaly small angle $\theta$. The contrast of our interferometer is limited by the noise on the cavity lock to $(R_{max}-R_{min})/(R_{max}+R_{min}) = 0.96$. This gives finite reflectivities to all the states, such that after a postselection on the reflected photon we are left with the components of both states $\ket{00}$ and $\ket{11}$ as error terms. For a given interferometer contrast, there is an optimal initial rotation angle that balances and minimizes the infidelities. The dependence of the maximal Bell state fidelity on the interferometer contrast and the initial angle is shown in Fig. \ref{fig1:angle} (left panel). Bell state fidelity is defined as the overlap between $\ket{\Psi^+}$ and the state created right after the photon reflection. For our interferometer contrast, we choose the initial angle of $0.3 \pi $ and expect the maximal fidelity of $0.76$. Given an initial angle and interferometer contrast, one can calculate the probability of having a photon reflected from the atom-cavity system, which for our parameters is $T_{int} = 0.1$.   

The protocol can be improved using a cavity with a smaller $R_{00}$, given that the maximal fidelity is intrinsically bounded by this value. Note that for our cavity linewidth the uncoupled reflectivity $R_{00}$ is equivalent to the empty cavity reflectivity. Fig. \ref{fig1:angle} (right panel) shows the dependence of the maximal fidelity and success probability on $R_{00}$. Assuming that the interferometer contrast stays the same, the fidelity can be improved to $0.9$ by reducing $R_{00}$ from $0.4$ (our parameter, black dot in the figure) to $0.1$. The tradeoff between the fidelity and the success probability is demonstrated by the opposite slopes of the solid and dotted curves in Fig. \ref{fig1:angle}. 

Finally, there are other avenues for improving the carving protocol beyond improving the interferometer contrast. For example, reflecting two gate photons separated by the coherent rotations could achieve a higher entanglement fidelity at the expense of a lower success rate, as demonstrated in~\cite{welte2017carving}. The success probability of the protocol can be increased by using a single photon source instead of an attenuated coherent pulse. In the case of an attenuated coherent pulse, the atomic coherence decays by a factor of $e^{-\lambda N_{sent}}$, where $\lambda = 4 (\kappa_{wg}/\kappa) C/(C+1)^2$ is the spontaneous emission probability of a strongly coupled atom~\cite{sorensen03}. We optimized $N_{sent}$ by measuring the retained coherence of single atoms after projecting them with coherent gate pulses and chose $N_{sent} = 0.35$ so that the spontaneous emission effect is negligible.

\subsection{Theoretical models for the entanglement protocol}

\subsubsection{Coherent model}

Ideally, the carving protocol for entanglement generation is fully coherent and results in a pure state. Here we model the effect of a coherent photon reflection and post-selection on detection. The model assumes that the photons reflected from different two-atom states differ by the reflection probability but are otherwise completely indistinguishable. Under this assumption, throughout the experiment the two atoms are in pure states, which are different superpositions of the Bell triplet states $\ket{\Phi^{\pm}}$ and $\ket{\Psi^{+}}$. The target entangled state from the main text is $\ket{\Phi^{+}}$, and we define the entanglement fidelity as the overlap $\mathcal{F}
=\bra{\Phi^+}\rho\ket{\Phi^+}$.  

We define the phase of our rotations such that a native rotation is over the axis x, so that after a rotation by angle $\theta$ we obtain $R_{0,\theta}\ket{00} = \ket{\Psi_\theta}$:
\begin{align} \label{eqn:psi_theta}
    \ket{\Psi_\theta}  =  & \cos^2\left(\frac{\theta}{2}\right) \ket{00} -  \sin^2\left(\frac{\theta}{2}\right) \ket{11} \nonumber \\
    & - i \sin\left(\frac{\theta}{2}\right)\cos\left(\frac{\theta}{2}\right) \left( \ket{01}+\ket{10}\right)
\end{align}

After preparing the state $\ket{\Psi_\theta}$, we reflect a photon and post-select on its detection. Given the reflection amplitudes for each state ($r_{00}$, $r_{01}$, $r_{10}$, $r_{11}$), we are left with 
\begin{align}
    \ket{\Psi^{cond}} = &\frac{1}{N} 
    ( r_{00}\cos^2\left(\frac{\theta}{2}\right) \ket{00} -
   r_{11}\sin^2\left(\frac{\theta}{2}\right) \ket{11} \nonumber \\
    & - i~r_{01}\sin\left(\frac{\theta}{2}\right)\cos\left(\frac{\theta}{2}\right) 
    \left( \ket{01}+\ket{10}\right) )
\end{align}
where $N$ is a normalization factor. If we relabel different state amplitudes as $\epsilon_0 = r_{00}\cos^2\left(\frac{\theta}{2}\right)/N$, $\epsilon_1 = r_{11}\sin^2\left(\frac{\theta}{2}\right)/N$, and $f = r_{01}\sqrt{2}\sin\left(\frac{\theta}{2}\right)\cos\left(\frac{\theta}{2}\right)/N$, we can rewrite the state $\ket{\Psi^{cond}}$ as
\begin{align}\label{eqn:state_decomposition}
    \ket{\Psi^{cond}} =&  ~\epsilon_0 \ket{00} - \epsilon_1 \ket{11} -i~f \ket{\Psi^+} \nonumber \\
                      =&   \left( \frac{\epsilon_0-\epsilon_1}{\sqrt{2}} \right) \ket{\Phi^+} + \left( \frac{\epsilon_0+\epsilon_1}{\sqrt{2}} \right) \ket{\Phi^-}
                      - i f \ket{\Psi^+}
\end{align}
Before analyzing the state, we apply $R_{0,\pi/2}$ and get the final state in the $ZZ$ basis 
\begin{align}\label{eqn:state_zz}
    \ket{\Psi_{\textrm{zz}}} = & ~R_{0,\pi/2} \otimes R_{0,\pi/2} \ket{\Psi^{cond}}\nonumber \\
                    = & -i\left( \frac{\epsilon_0-\epsilon_1}{\sqrt{2}} \right) \ket{\Psi^+} + \left( \frac{\epsilon_0+\epsilon_1}{\sqrt{2}} \right) \ket{\Phi^-}
                      - f \ket{\Phi^+}
\end{align}
Based on Eqn. \ref{eqn:state_zz}, the entanglement fidelity is given by $\mathcal{F} = |f|^2$. The idea of the "cavity carving" protocol is to maximize the amplitude $f$. This can be done by 1) having ($r_{00} \ll r_{01} \approx r_{11}$), which would result in a small $\epsilon_0$, and 2) using a small initial rotation angle $\theta$, which would result in a small $\epsilon_1$.

From Eqn. \ref{eqn:state_zz} we can determine the population in the four basis states:
\begin{align}\label{eqn:populations}
    P_{00} = &~\frac{1}{2} \left |\frac{\epsilon_0+\epsilon_1}{\sqrt{2}} - f \right|^2 \\
    P_{11} = &~\frac{1}{2} \left |\frac{\epsilon_0+\epsilon_1}{\sqrt{2}} + f \right|^2 \\
    P_{01} = &~P_{10} = \frac{1}{4} \left |\epsilon_0-\epsilon_1\right|^2 
\end{align}

Note that finite values of $\epsilon_0$ and $\epsilon_1$ cause the populations in $\ket{00}$ and $\ket{11}$ to be unbalanced (as seen in Fig.~3B).

\subsubsection{Mixed model}

\begin{figure*}
\centering
\includegraphics[width=1.4\columnwidth]{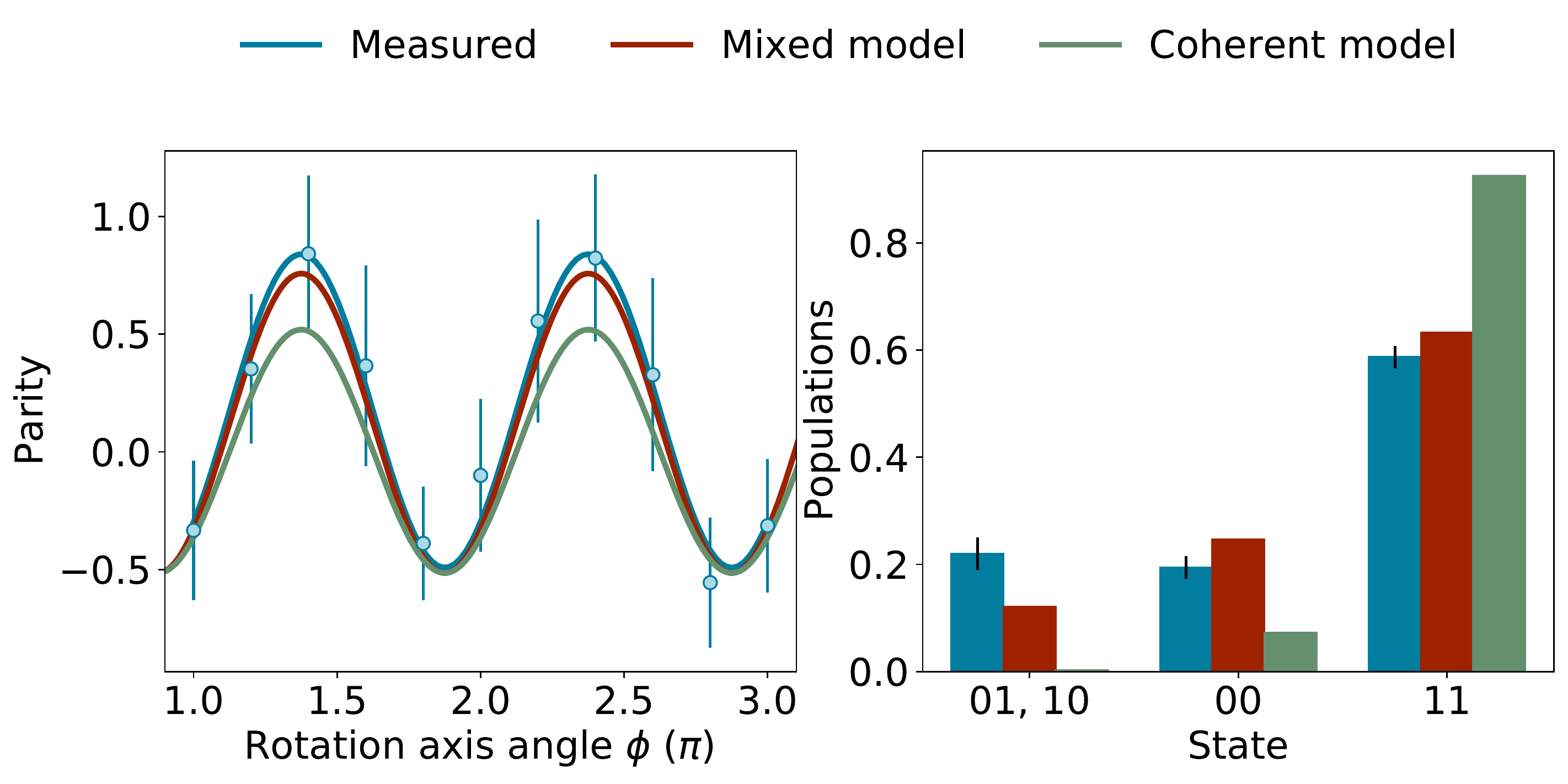}
\caption{Comparison of the experimental measurements (blue), mixed model predictions (red), and coherent model predictions (green) for the parity measurement (left panel) and the $ZZ$ basis populations (right panel).}
\label{fig:theory}
\end{figure*}

The coherent model above assumes that the post-selection on the reflected photon leaves the system in a pure state. However, spontaneous scattering and temporal mismatch can both reveal the which-path information of photons reflected from $\ket{00}$. When the photons reflected from the uncoupled and coupled states are distinguishable, the system is left in a mixed state. Spontaneous scattering from the gate photons in the coherent pulses containing more than one photon leave the atom in a different hyperfine level within the $F=2$ manifold. In addition, while the response from the state $\ket{0}$ is instantaneous, the bandwidth of the response from the state  $\ket{1}$ is $C\gamma$ and may lead to a temporal mismatch. The atom samples a distribution of cooperativities due to thermal motion, which for $C\sim1$ results in a $26~$ns delay in the atomic response. This is longer than the pulse duration of $20~$ns (FWHM) and would make the photons distinguishable.

We can model distinguishability by assuming that a photon reflection corresponds to the following projections:

\begin{align}
M_u = \ket{00}\bra{00} \\
M_c = \unit - M_u
\end{align} 

If the probability that a detected photon was reflected from $\ket{00}$ is $p_u$ and from the coupled manifold is $p_c = 1 - p_u$, post-selection on a photon detection projects the system into

\begin{align}
\rho^\prime = p_u M_u \rho M_u^\dagger + p_c M_c \rho M_c^\dagger
\end{align}

where $\rho$ is the density matrix of the state before reflection, $\rho = \ket{\Psi_\theta} \bra{\Psi_\theta} $ (Eqn. \ref{eqn:psi_theta}). The probability $p_u$ can be calculated as the ratio of effective reflectivities of the uncoupled and coupled states (assuming that $R_{10} \approx R_{11}$):

\begin{align}
p_u &= \frac{R_{00}}{R_{00}+R_{01}} 
\end{align}

When using an interferometer, $R_{00}$ and $R_{01}$ would correspond to reflectivities with the interferometer engaged. For our measured interferometer contrast of $0.96$, $p_u = 0.087$. 

The mixed model differs from the coherent model in that the reflection post-selection destroys the initial coherences between $\ket{00}$ and the coupled states: $\rho_{00,01}$, $\rho_{00,10}$, $\rho_{00,11}$. When the state is further rotated and analyzed, such as in the populations and parity oscillation measurement, this will result in different theoretical predictions. However, we note that the entanglement fidelity prediction of the two models is identical, since the entanglement relies on the reflection preserving the initial coherence between $\ket{01}$ and $\ket{10}$ ($\rho_{01,10}$) after the gate pulse, which happens in both models. We confirm this statement numerically, by calculating the overlap with $\ket{\Phi^+}$ with different models, and arriving at the same values. 

Given that both the coherent and the mixed model give predictions for the state populations and measured parity, we here compare them with the measured data. Figure \ref{fig:theory} shows the experimental measurement (blue), predictions from the coherent model (green), and mixed model (red) for the parity oscillation (left panel) and $ZZ$ basis populations (right panel). The mixed model agrees better with the measurements in both datasets.

\subsection{Coherence properties of atoms next to the photonic crystal cavity}

The qubit relaxation time $T_1$ in our system is limited by contributions of various scattering processes. The tweezer depth next to the PCC ($32$ MHz) would limit $T_1$ to $60 $ ms. However, with the cavity tuned into resonance with the atom, we observe additional scattering that significantly reduces $T_1$ to $1.2$ ms. With the cavity tuned $14\kappa$ away from the atomic resonance, we recover the expected $T_1$. For the experiments that require long coherence, such as for atomic state transport, we detune the cavity right after the gate by jumping the cavity lock in sequence. The source of the resonant scattering remains under investigation.   
 
\begin{figure}[h!]
\centering
\includegraphics[width=0.7\columnwidth]{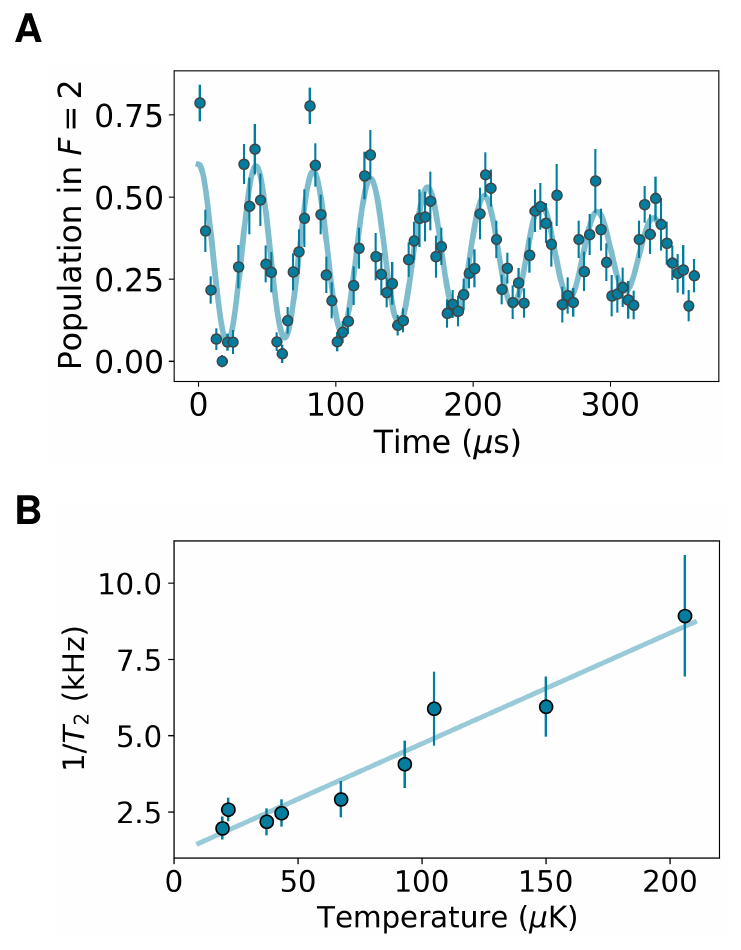}
\caption{ Temperature-limited Ramsey decay. 
\textbf{(A)} Ramsey decay of an atom next to the PCC fitted using the model described in \cite{dephasing_meschede}, giving $T_2 = 0.35(4)$ ms.
\textbf{(B)} Linear dependence of the Ramsey dephasing rate on the temperature. 
}
\label{fig:ramseys}
\end{figure}

Magnetic-field-insensitive qubit encoding minimizes the influence from vector light shifts, a near-field effect intrinsic to tightly-focused light and further exacerbated by the standing wave potential next to the PCC \cite{thompson2012raman}. 
The spin dephasing is characterized by the transverse decay time $T_2$ and is primarily limited by thermal motion leading to varying differential AC Stark shifts. 
This dephasing can be decomposed into the reversible ($T_2^*$) and the irreversible ($T_2'$) component as $1/T_2 = 1/T_2^\prime + 1/T_2^*$. Shown in Fig. \ref{fig:ramseys}A is a Ramsey decay measurement $\left( \pi/2 - \tau - \pi/2 \right)$ of an atom next to the PCC, giving a $T_2 = 0.35(4)$ ms. As discussed in \cite{dephasing_meschede,dephasing_fiber}, the main source of reversible dephasing in optical dipole traps is atomic motion coupled with a finite differential light shift. If we assume that the energy distribution of atoms in the trap obeys the Boltzmann distribution with temperature $T$, the reversible dephasing time $T_2^*$ can be expressed as $T_2^* = 2\hbar/(\zeta k_B T)$ \cite{dephasing_meschede}. The factor $\zeta$ depends on the dipole trap wavelength and the atomic energy spectrum. We experimentally confirm the linear dependence of $1/T_2$ on the temperature and  measure $\zeta = 5.4 \times 10^{-4}$ (Fig. \ref{fig:ramseys}B). The dephasing induced by a constant temperature can be reversed with a spin echo sequence.  

\begin{figure}[h!]
\centering
\includegraphics[width=0.8\columnwidth]{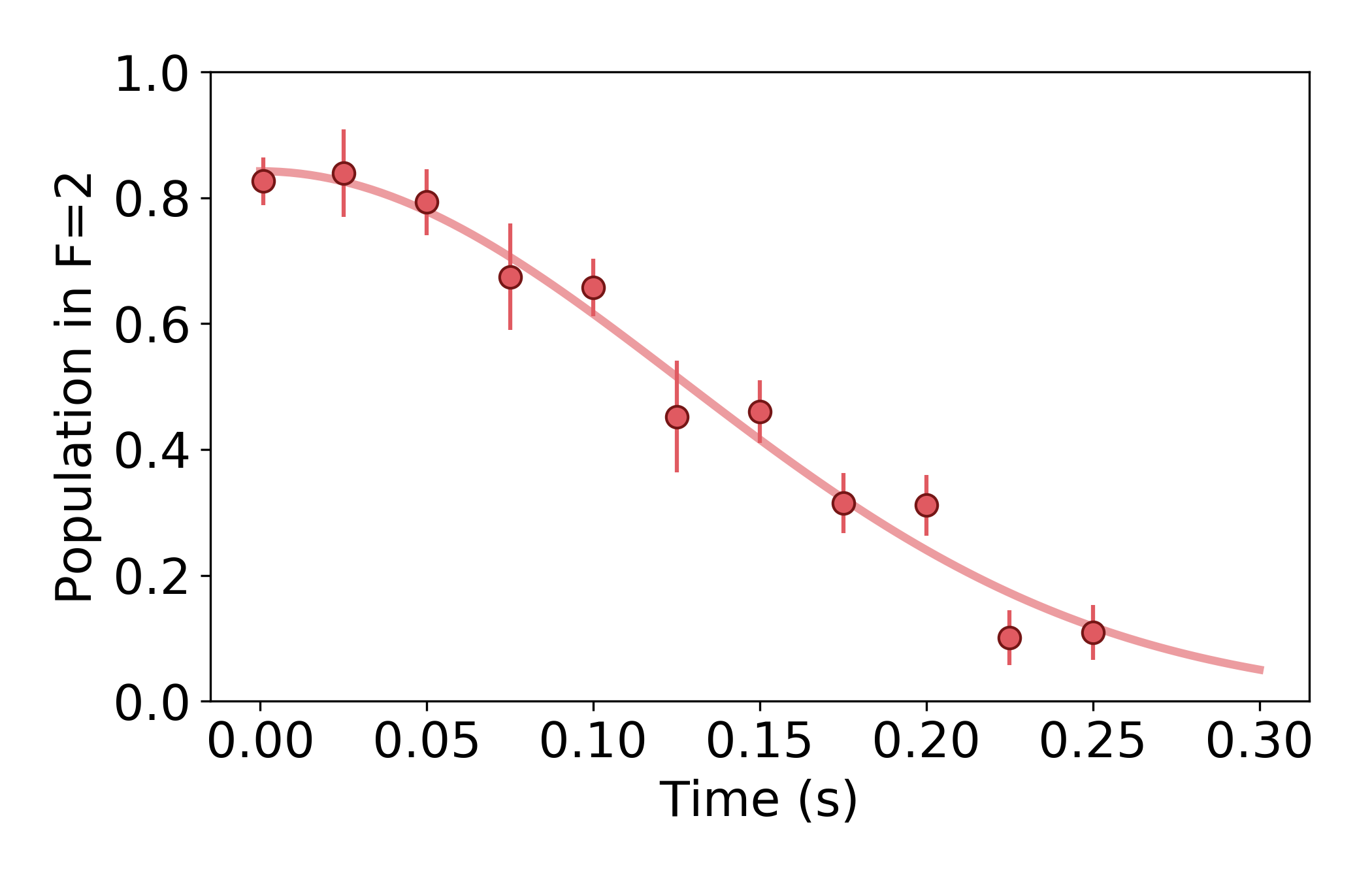}
\caption{ Measured lifetime of $180(7)$~ms at a trap depth of $2.7$~mK next to the PCC. }
\label{fig:lifetime}
\end{figure}

Using the spin echo pulse sequence $\left( \pi/2 - \tau - \pi - \tau - \pi/2 \right)$, we measure $T_2^\prime = 9.7(8)$ ms in free-space and $2.1(1)$ ms at the cavity. The difference cannot be explained with the sources common to the free space and the PCC such as fluctuating magnetic fields, microwave power and pulse duration, and we therefore attribute it to extra heating at the cavity. Any heating process would translate to irreversible dephasing, in the same way constant temperature causes reversible dephasing. At the cavity, we measure a lifetime of $180(7)$ ms at a trap depth of $2.7$ mK (Fig. \ref{fig:lifetime}), giving a heating rate of $14.6$ mK/s, an order of magnitude higher than in free space. The irreversible dephasing time depends on the heating rate as $T_2' \propto (dT/dt)^{-1/2}$ \cite{dephasing_meschede}.   

Two main sources of heating we consider are  trap potential fluctuations and atomic recoil due to scattering. To estimate the heating rate due to atomic recoil, we use the scattering rate given by $1/T_1$ and obtain $dT/dt = 4.1 \cdot 10^{-3} $ mK/s (cavity out of lock) and $0.15$ mK/s (cavity in lock). Both estimates differ by orders of magnitude from the lifetime-deduced heating rate. 
Therefore, we speculate that the contributions to additional heating near the nanophotonic cavity include trap fluctuations due to mechanical vibrations of the nanostructure~\cite{hummer19} and higher sensitivity of the axial trap frequency to the fast intensity noise~\cite{Savard97}.

\subsection{Phase accumulation while moving}
\subsubsection{Dipole potential for different trap configurations}
The potential energy of a two-level system in a detuned optical field in the rotating-wave approximation is given by: 
$$U(r) = \frac{3 \pi c^2 \gamma}{2 \omega_0^3 \Delta} I(r)$$
where $\Delta$ is the field detuning, $\gamma$ the spontaneous emission rate, and $I(r)$ the intensity of the field \cite{dt_review}. When the field is red-detuned ($\Delta<0$), the atoms are attracted to the maximum of the field intensity $I(r)$. Based on the different configurations of the laser beams, $I(r)$ can have different spatial dependence, resulting in different effective trap frequencies. In our experiment, we change the trapping configuration from a focused Gaussian beam to a standing wave. Here we will explore how the potential changes for those two configurations. 

A focused Gaussian beam creates a potential
\begin{equation}
    U(r,z)= U_0 \exp{\frac{-2 r^2}{w(z)^2}}
\end{equation}
where $w(z) = w_0 \sqrt{1+\frac{z^2}{z_r^2}}$, $w_0$ is the beam waist, and $z_r = \pi w_0^2/\lambda$ is the Rayleigh range. Expanding this expression around $r,z = 0$ gives
\begin{equation}
    U(r,z) \approx -U_0 \left( 1 - 2 \left(\frac{r}{w_0}\right)^2 - \left(\frac{z}{z_r}\right)^2 \right)
\end{equation}
which can be rewritten as a harmonic potential in the radial and axial direction
\begin{equation}
     U(r,z) \approx - U_0 + \frac{1}{2}m \omega_r^2 r^2 + \frac{1}{2}m \omega_z^2
\end{equation}
with the oscillator frequencies $\omega_r = \sqrt{\frac{4 U_0}{m w_0^2}}$ and $\omega_z = \sqrt{\frac{2 U_0}{m z_r^2}}$. When the atoms are near the nanophotonic cavity, part of the tweezer beam is reflected back and interferes with the forward-propagating beam. The tweezer then forms a standing wave potential of the form
\begin{equation}
    U(r,z)^{'} = -\alpha U_0 \cos(kz)^2 \left( 1 - 2 \left(\frac{r}{w_0}\right)^2 - \left(\frac{z}{z_r}\right)^2 \right)
\end{equation}
where $\alpha$ is the interference contrast (measured to be $\alpha=1.2$ in our experiment), and $k = 2\pi/\lambda_0$. Expanding this expression and rewriting it as a harmonic potential gives the oscillator frequencies $\omega_r^{\prime} = \sqrt{\frac{4 \alpha U_0}{m w_0^2}}$ and $\omega_z^{\prime} = \sqrt{\frac{2 \alpha U_0}{m z_{\lambda}^2}}$, where $z_{\lambda} = \lambda_0/(2\pi)$. Therefore, moving from the focused beam to a standing wave changes the oscillator frequencies from $\omega_r \rightarrow \omega_r^{\prime}$ and $\omega_z \rightarrow \omega_z^{\prime}$.

\subsubsection{Differential light shift during the transport}
The main contribution to the shift of the qubit resonance for the magnetic-field-insensitive states is the differential light shift. For a given maximal light shift $U_0$, the differential light shift $\delta_0$ is given by
\begin{equation}
    \hbar \delta_0 = \zeta (-U_0 + \overline{E_{p}})
    \label{eq:si1}
\end{equation}
where $\zeta \propto \omega_{HFS}/\Delta$ ($\zeta = 5.4 \times 10^{-4}$ for our system), and $\overline{E_{p}}$ is the average potential energy of the atom in the trap. From the virial theorem, $\overline{E_{p}} = \overline{E}/2$. For the three-dimensional harmonic oscillator, the average total energy is given by 
\begin{align}
    \overline{E} =& \sum_{i = x,y,z} \hbar \omega_i 
    \left(\frac{1}{2} + \frac{1}{\exp{\frac{\hbar \omega_i}{k_b T}} - 1} \right) \nonumber \\
    =&  \sum_{i = x,y,z} \hbar \omega_i  \left(\frac{1}{2} + n_i \right)
    \label{eq:si2}
\end{align}
where $T$ is the temperature of the atom, and $n_i$ is the average occupation number in the direction $i$. Combining the Eqn. \ref{eq:si1} and \ref{eq:si2}, we arrive at the expression for the differential light shift of a moving atom:
\begin{align}\label{eq:si3}
    \delta_0(t) = -\zeta \frac{U_0(t)}{\hbar} + 
    \frac{\zeta}{2} 
    \Bigg( 2&\omega_r(t) \left( \frac{1}{2} + n_r(t)\right) +   \nonumber \\ 
    & \omega_z(t) \left( \frac{1}{2} + n_z(t)\right) \Bigg)
\end{align}
As seen in the previous section, $U_0(t)$, $\omega_r(t)$ and $\omega_z(t)$ necessarily vary in time as the trap potential changes. The average occupation number $n_i(t)$ depends on whether the atom makes motional state transitions during the transport. If the trap deformation is adiabatic, the atom does not change the motional energy level and $n_i =  const$. This is equivalent to $\omega(t)/T(t) = const$, which would imply an increase in temperature as the trap frequency increases (adiabatic heating) and vice-versa (adiabatic cooling)\cite{Tuchendler08}. The entire time dependence of $\delta_0(t)$ would then be in $U_0(t)$, $\omega_r(t)$ and $\omega_z(t)$. If the atom does make motional state transitions because of the noise in the trapping potential, $\delta_0(t)$ will stochastically vary and cause qubit dephasing. 

\subsubsection{Phase accumulation and cancellation}

Given that $\delta_0$ varies in time as shown in the Eqn. \ref{eq:si3}, a moving atom will acquire a phase shift $\phi(T) = \int_0^{T} \delta_0(t) dt$. If the detuning ramp during transport is the same for every atom, i.e. if the acquired phase is constant, it can be corrected for with a local $\sigma_z$ gate. 

\begin{figure*}
\centering
\includegraphics[width=1.4\columnwidth]{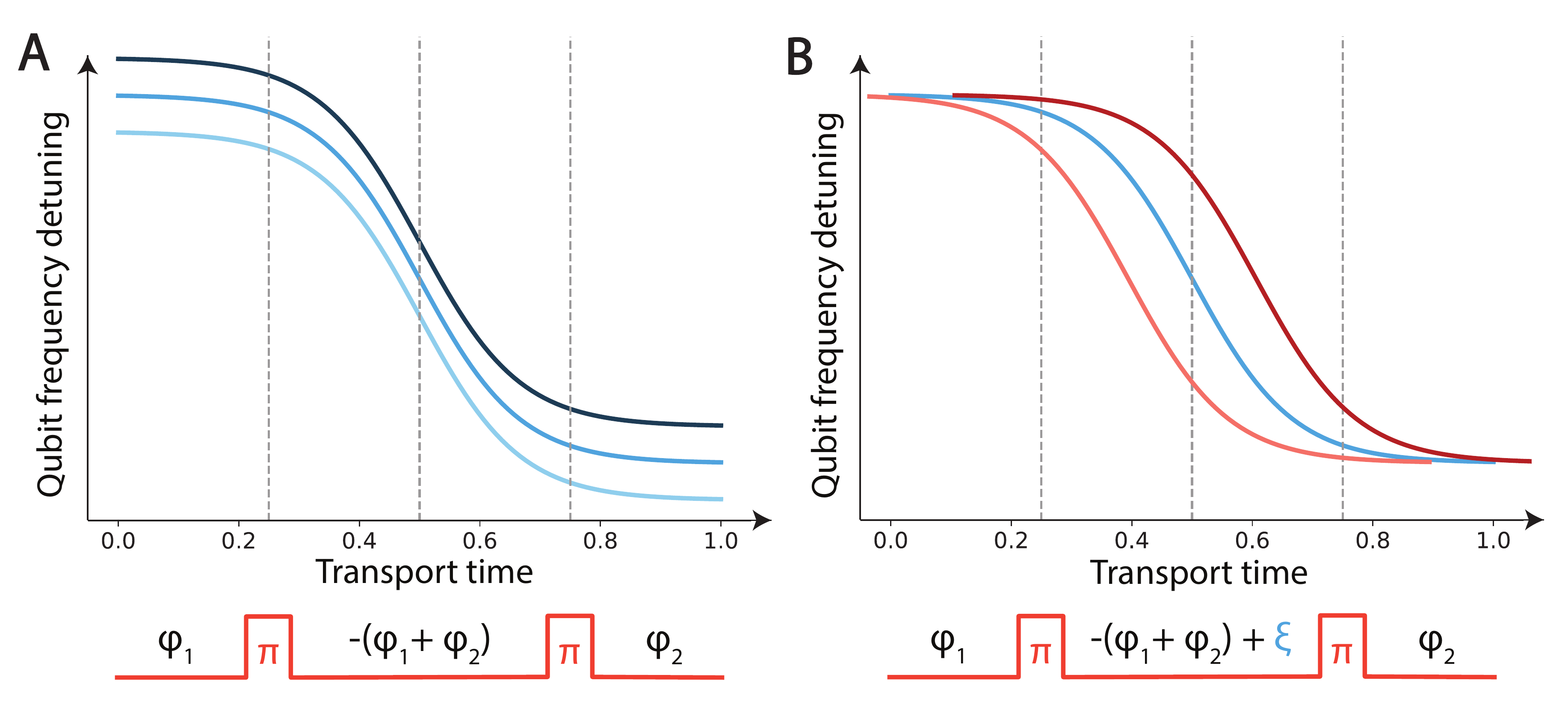}
\caption{Sketches of how variations in detuning can cause dephasing. \textbf{(A)} Variations in the overall qubit frequency offset (vertical variations), caused by atom's finite temperature, can be echoed out with an even number of CP pulses. \textbf{(B)}  Variations in the moving trajectory (horizontal variations), caused by trap pointing instabilities, cannot be echoed out exactly and contribute to extra dephasing.}
\label{fig2:detuning_ramp}
\end{figure*}

We apply a voltage ramp $V(t) \propto \tanh(t)$ to the scanning mirrors to move the atoms. If the atoms were to follow the mirror position instantaneously, the detuning ramp would have the shape as plotted in Fig. \ref{fig2:detuning_ramp}. 
The ramp is symmetric with respect to $T/2$. An even number of $\pi$ pulses in a Carr-Purcell (CP: $\pi/2 - (\tau - \pi - \tau)^N - \pi/2$) sequence would cancel out the acquired phase. The statistical variations in the total energy in one experimental run that come from a finite temperature would result in a variation in the offset of the detuning ramp (Fig \ref{fig2:detuning_ramp}A), which is the same as the variation that limits the free precession time in a Ramsey sequence. Analogously to echoing out the phase with one spin echo in a static case, an even number of spin echo pulses would echo it out in the symmetrically changing case. Importantly, this configuration would echo out the phase variation due to the finite temperature, thus reverting qubit dephasing. If the trap pointing varies in time such that the detuning symmetry axis varies (Fig \ref{fig2:detuning_ramp}B), the symmetric echo would not correct for the temperature-induced variations, and the moving atoms would have an extra dephasing channel. The pointing variations are minimized by tracking the cavity position periodically. To move the traps, we use different scanning mirrors (trap A: Cambridge Technology 8310K, trap B: Physik Instrumente S-330.4SL). We have observed differences in coherence properties between the two traps after transport, which we attribute to larger fluctuations in trap A. 

In order to distribute the pulses symmetrically, we map out the detuning profile during the transport. We probe each of the atoms by starting in the state $\ket{0}$ and applying a detuned $\pi$ pulse at a variable point during the transport. The population transfer to $\ket{1}$ will depend on the qubit frequency at the time of the $\pi$ pulse according to the Rabi lineshape. We measure and account for the delay in response for each galvanometer mirror and calibrate their input voltage to symmetrize the ramp with respect to the moving time, as shown in Fig.~2B. Furthermore, the bandwidth of the trap B mirror limits the step response time to $\sim 650~\mu$s, and we program the ramp of trap A to match. This is the time over which the CP sequence is applied as described in the main text. 

\begin{figure*}
\centering
\includegraphics[width=1.4\columnwidth]{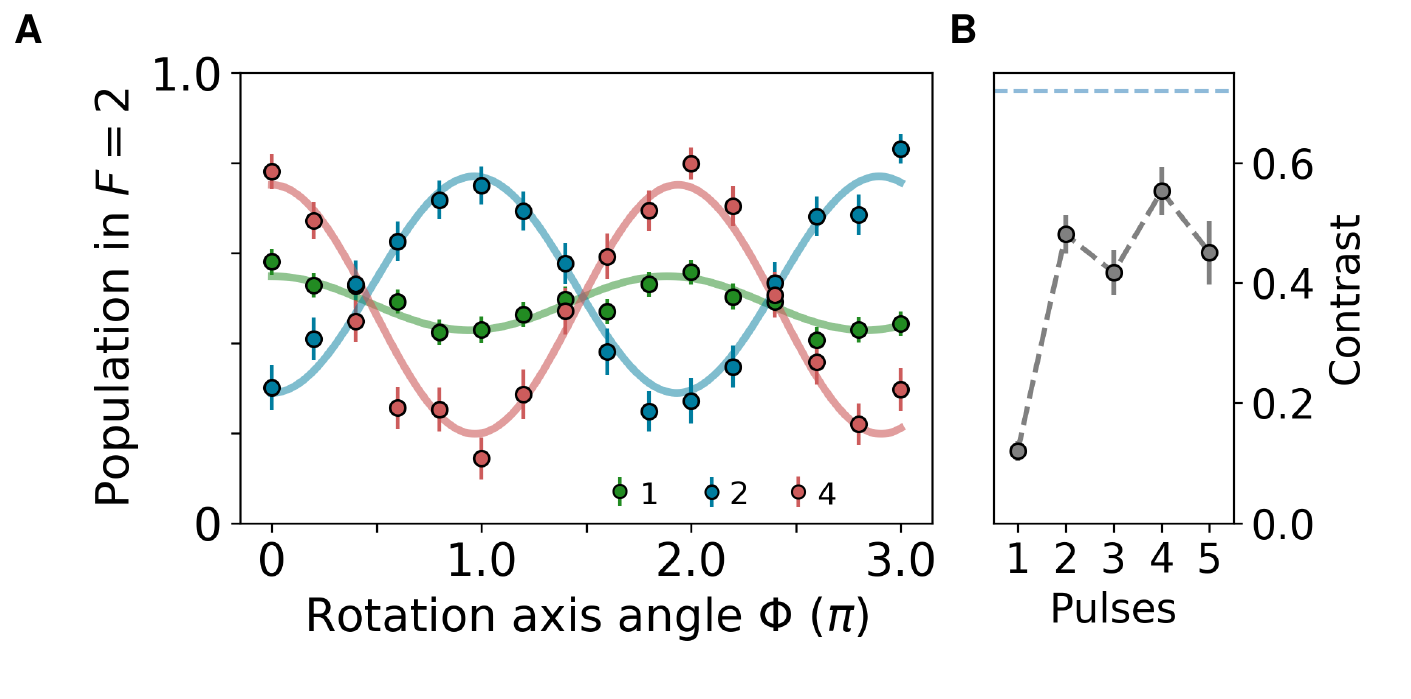}
\caption{Optimizing the number of pulses in the CP sequence for a moving atom. 
\textbf{(A)} CP sequences of $N = (1,2,4)$ pulses that begin with an atom at the cavity and end with an atom in free space. Decoupling pulses are applied in parallel with the transport.
\textbf{(B)} Fitted contrast of the CP sequences versus the number of CP pulses. 
}
\label{fig:cpmg}
\end{figure*}

We optimize the number of pulses in the CP sequence by preparing an atom in a superposition state at the PCC, moving it away while applying CP, and studying the remaining coherence contrast after a variable number of pulses, as shown in Fig. \ref{fig:cpmg}. The even number of pulses performs better than the odd one, as expected, and we find the optimal number to be four for our system parameters. Higher frequency noise components could be addressed by having faster $\pi$ pulses (in our system $T_{\pi} = 20~\mu$s). 
In the main text, we report on the retained coherence of the trap B after moving to be $0.88(5)$~(Fig. 2C). For a stationary atom, $N = 4$ CP sequence gives $T_2^\prime = 4.9(4)$ ms ($17(2)$ ms) for atoms at (away from) the PCC. If we assume that half of the move time (325~$\mu$s) is spent at the cavity and half in free space, we expect to retain $0.92$ of the coherence for a $T_2^\prime$-limited transport, which is within the uncertainty of our measured result. From here we deduce that the used decoupling sequence is effective at canceling out the moving-induced dephasing.

\subsection{Modeling losses upon loading into the cavity near-field}
Integration of our platform with a free space atom array would require low losses upon loading into the cavity near-field. Here, we theoretically model what those losses depend on and derive requirements for lossless loading. We study the effects of transferring the atom from the tweezer potential to the optical lattice potential in the near-field of the cavity, resulting from the reflection of the tweezer off of the cavity.

We numerically study classical trajectories of a single particle in an evolving gaussian potential and observe that an adiabatic increase in the trap frequency introduces heating that ultimately leads to losses, the reverse process of adiabatic cooling (see Section 8.1)~\cite{Tuchendler08,alt2003diabatic}.
The thermal loss fraction can be determined by averaging trajectories over the phase space weighted by the Boltzmann factor. As there are two relevant axes that transform upon tweezer arrival 
($\omega_r$, $\omega_z$) $\rightarrow$ ($\omega_r^{\prime}$, $\omega_z^{\prime}$) \cite{thompson2013couplingpc}, we generate the loss fraction as a function of the temperature for the principal axes separately and deduce the survival from
the joint probability. Our model assumes separability between the two principal axes,
continuity of the trap deformation, and no extra heating caused by the trap rotation. The losses as functions of temperature are generated with parameters extracted from independent trap frequency measurements.

\begin{figure}
\centering
\includegraphics[width=1.0\columnwidth]{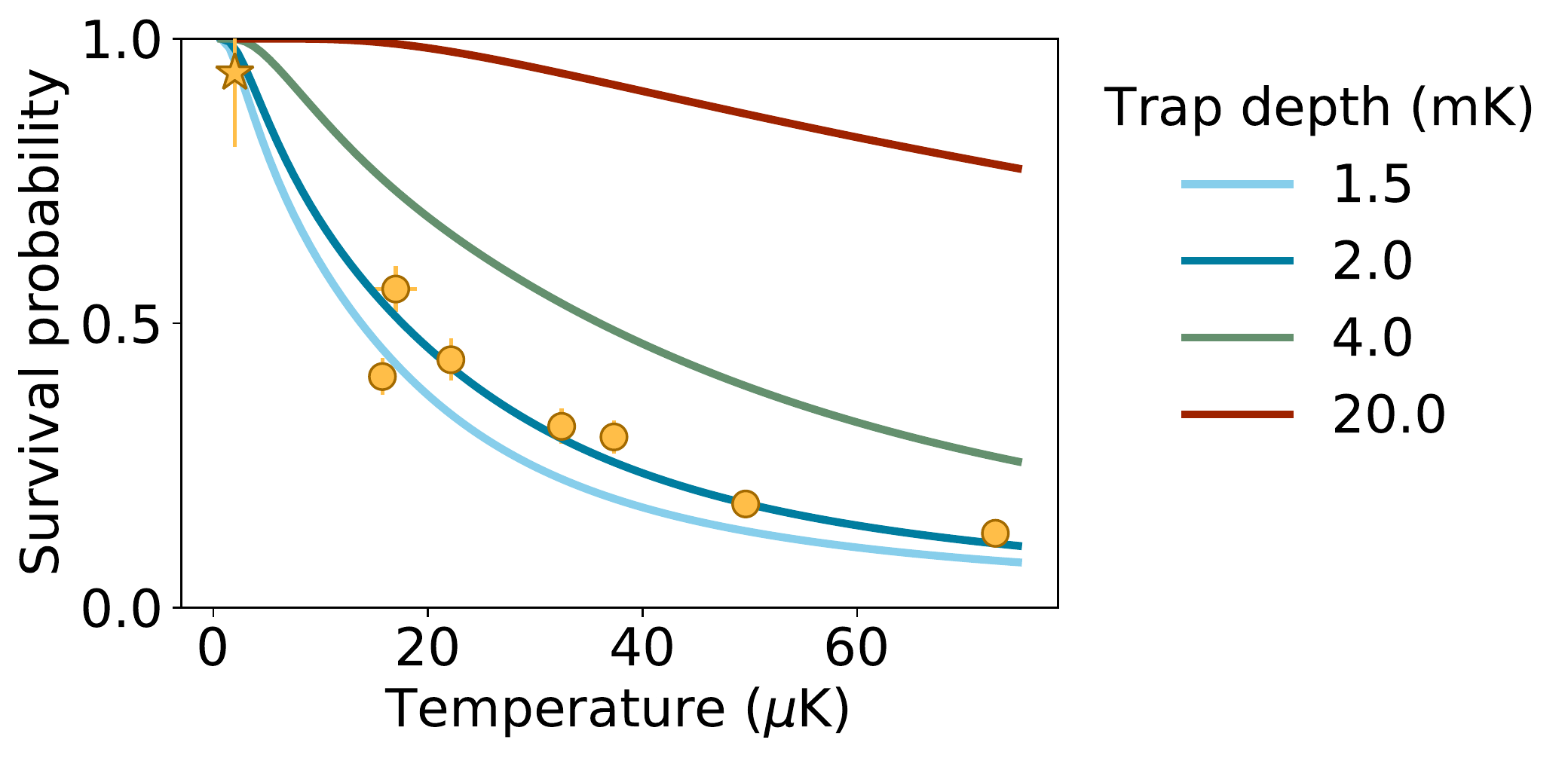}
\caption{The effects of temperature and trap depth on the survival probability of an atom loading into the cavity near-field. The theoretical curves are generated by
averaging over classical trajectories weighted by the Boltzmann factor
with no free parameters. Our experimental observations (yellow circles) are taken at a trap depth of $1.6~$mK.
Previous work on our platform with Raman sideband cooling before transport observed $0.94(6)$ survival probability (yellow star) \cite{thompson2013couplingpc}.}
\label{fig:loss}
\end{figure}

Figure \ref{fig:loss} shows the theoretical predictions for the loading survival probability as a function of the initial atomic temperature and trap depth. 
Our survival measurements without cooling at different initial temperatures and a trap depth~$U_0 = k_B \times 1.6$~mK are marked with yellow circles. 
The initial temperature is varied by changing the free space PGC parameters, and the temperature values are extracted from drop-recapture measurements. 
Our observations follow the same trend as the theoretical model.

Our numerical results suggest several ways in which the arrival losses can be remedied.
Since loading into the near-field causes heating, applying cooling in parallel with the transport would counter some of it. We employ this method in this work by performing polarization gradient cooling (PGC) in parallel with the transport and achieve a survival probability of $0.8$. This method is not suitable if the atomic internal states need to be preserved. In that case, other loading improvements are possible.
The most immediate one is to cool to the motional ground state prior to loading, which would also improve coherence properties of atoms by reducing the sensitivity to light shifts. 
Previous work on our platform utilizing Raman sideband cooling prior to transport (Fig. \ref{fig:loss}, yellow star)
has achieved a survival probability of 0.94(6) at $T = 2~\mu$K
and a trap depth~$U_0 = k_B \times 2.1$~mK \cite{thompson2013couplingpc}.
Without additional cooling, ramping the trap depth up could increase the survival probability, but at the expense of reduced coherence due to higher light shifts. 
Finally, in our model the losses come from the change in trap frequencies, and applying additional light shift during transport could compensate for that change.  This could be achieved with additional trapping beams perpendicular to the current traps that would increase the initial motional frequencies, and that would be ramped down as the traps approached the cavity.

Based on the model presented here, there should be no loss upon departure from the cavity since the trap frequencies decrease. Yet, at a trap depth of~$U_0 = k_B \times 1$~mK we observe a $15\%$ loss. Further investigation is needed to model the loss at low trap depths, but we observe a lossless transport at a trap depth of~$U_0 = k_B \times 1.6$~mK, making our current system already suitable for protocols that rely on one-way transport.


\end{document}